\documentclass[aps,prb,twocolumn,amsmath,amssymb,superscriptaddress,longbibliography]{revtex4-2}
\usepackage{graphicx}
\usepackage{float}
\usepackage{multirow}
\usepackage{natbib,twoopt}
\usepackage[breaklinks=true]{hyperref}
\usepackage{xcolor}
\usepackage{amssymb}
\usepackage{stackengine,graphicx}
\usepackage{amsmath}
\usepackage{hyperref}
\usepackage{color}
\usepackage{stackengine}
\usepackage{subfigure}

\usepackage{verbatim}
\usepackage{stmaryrd} 

\newcommand{\divv}{\mathop{\rm div}\nolimits}

\newcommand{\eps}{\varepsilon}
\newcommand{\om}{\omega}
\newcommand{\kap}{\kappa}
\newcommand{\bas}{{\bf e}_z}

\graphicspath{{Figures/}}
\renewcommand{\v}[1]{\mathbf{#1}}
\let\Im\undefined
\DeclareMathOperator{\Im}{Im}
\let\Re\undefined
\DeclareMathOperator{\Re}{Re}


\begin{document}

\title{Effective chiral response of anisotropic multilayered metamaterials}

\author{Konstantin Baranov}
\email{These authors have contributed equally to this work}
\affiliation{School of Physics and Engineering, ITMO University, Saint  Petersburg 197101, Russia}

\author{Dmitry Vagin}
\email{These authors have contributed equally to this work}
\affiliation{School of Physics and Engineering, ITMO University, Saint  Petersburg 197101, Russia}

\author{Maxim A. Gorlach}
\email{m.gorlach@metalab.ifmo.ru}
\affiliation{School of Physics and Engineering, ITMO University, Saint  Petersburg 197101, Russia}

\begin{abstract}
We consider a multilayered metamaterial composed of anisotropic layers with in-plane anisotropy axes periodically rotated with respect to each other. Due to the breaking of inversion symmetry, the structure develops an effective chirality. We derive the effective chirality tensor of the metamaterial describing both normal and oblique incidence scenarios. We also identify the arrangement of the layers maximizing chiral response for the fixed lattice period, wavelength and anisotropy of the layers.
\end{abstract}

\maketitle

\section{Introduction}

One of the most salient features of metamaterials is the emergence of effective properties beyond those of their constituent elements. While elementary building blocks of the structure are ordinary metals or dielectrics, their collective response to the electromagnetic excitation can be quite nontrivial. Besides celebrated negative effective permittivity and permeability~\cite{Veselago1968,Eleftheriades2005,Marques2008}, emergent properties include various forms of electromagnetic nonreciprocity~\cite{Belotelov2007,Asadchy2018}, nonlocality~\cite{Belov2003,Orlov2011} and even effective axion fields~\cite{Shaposhnikov2023,Asadchy2024}.

One of such useful properties is  chirality which is broadly defined as impossibility to superimpose object with its mirror image. Specifically, we focus on {\it electromagnetic chirality} that couples electric and magnetic responses of the medium, ensures different interaction of the medium with left- and right-hand circularly polarized waves~\cite{Semnani2020,Voronin2022,Baranov2023} and originates due to the breaking of inversion symmetry. While different measures of chirality are discussed in the literature~\cite{Tang2010,Rockstuhl2016,Gorkunov2024}, here we introduce it as a tensor $\hat{\kappa}$ arising in the bulk material equations~\cite{Serdyukov2001}
\begin{equation}
    \label{eq:bianis_form}
        \begin{cases} \v{D}=\hat{\varepsilon}\v{E}+i\hat{\kappa}\v{H},\\
    \v{B}=-i\hat{\kappa}^\mathsf{T}\v{E}+\hat{\mu}\v{H},
    \end{cases}
\end{equation}
where $\hat{\eps}$ and $\hat{\mu}$ are the conventional permittivity and permeability.

Chirality is ubiquitous in organic materials, where it stems from the non-centrosymmetric nature of the constituent molecules and can hardly be tuned. At the classical level, this physics is captured by the Born-Kuhn model~\cite{Yin2013}; quantum-mechanical description is also available~\cite{Barron}.

At the same time, chirality can also be configurational arising due to the proper non-symmetric arrangement of centrosymmetric elements~\cite{Vetrov2020,Rebholz2024}. This is the case for cholesteric liquid crystals utilized for LC displays technology~\cite{deGennes1993,Khoo2022}. More recently, helical homostructures have been realized by stacking biaxial van der Waals crystals exhibiting large permittivity up to 10 along with strong anisotropy~\cite{Voronin2024}. These experimental advances open a route to tailor chiral response of optical nanostructures. 

However, the detailed theoretical understanding of emergent chirality is currently lacking. As previous studies focused on the normal incidence scenario and extracted only single component of chirality~\cite{Vetrov2020,Voronin2024}, full tensor $\hat{\kappa}$ remained unknown. Furthermore, emergent chirality depends on the specific way how anisotropic layers are rotated with respect to each other. While uniform rotation from layer to layer is the most straightforward option, other arrangements are also possible and could potentially bring richer phenomena.

In this Article, we aim to fill this gap and develop an effective medium description of a metamaterial composed of anisotropic layers with in-plane anisotropy axes rotated with respect to each other. Building on homogenization theory~\cite{Silveirinha2007,Gorlach2020}, we retrieve the full set of the effective material parameters capturing normal and oblique incidence geometry. The obtained results are general and encompass the entire family of structures with arbitrary periodic law of anisotropy axis rotation. In turn, this allows us to maximize the effective chirality depending on the layer arrangement for the fixed period, wavelength and layer permittivities.

The rest of the Article is organized as follows. In Sec.~\ref{sec:Symmetry} we examine the symmetry constraints on the structure of chirality tensor. We outline our approach for the normal incidence scenario in Sec.~\ref{sec:NormalInc} and compare the key predictions with the rigorous transfer matrix method in Sec.~\ref{sec:Validation}. Having validated the analytical results, in Sec.~\ref{sec:Maximization} we identify the arrangement of the layers favoring maximal chiral response. Eventually in Sec.~\ref{sec:ObliqueInc} we reconstruct full chirality tensor featuring somewhat counter-intuitive structure. Finally, we conclude by Sec.~\ref{sec:Discussion} summarizing our results and outlining future perspectives.


\section{Symmetry analysis}\label{sec:Symmetry}

\begin{figure}
    \centering
    \includegraphics[width=0.9\columnwidth]{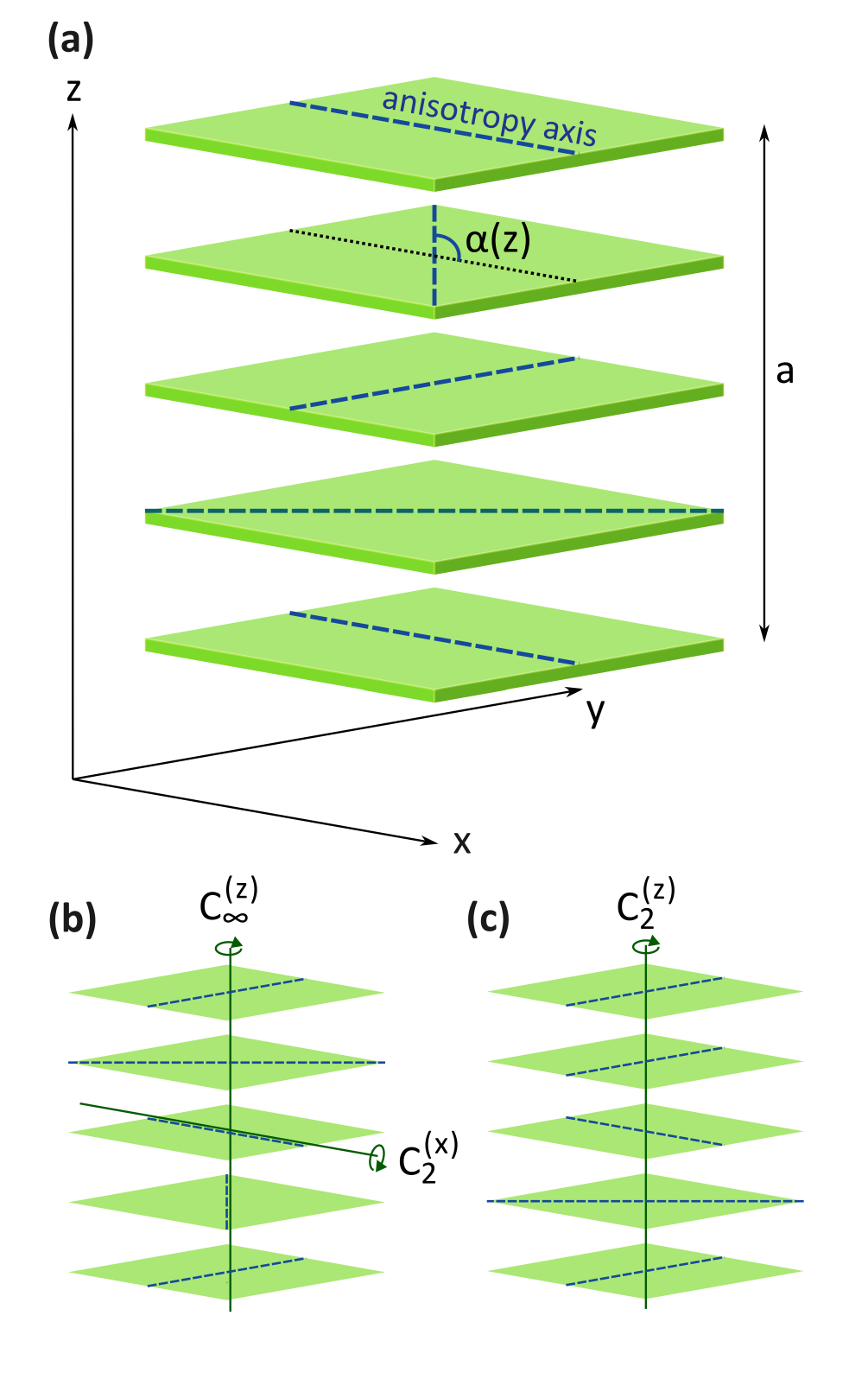}
    \caption{(a) A sketch of the multilayered structure composed of anisotropic layers rotated relative to each other. (b) Point symmetry axes in case of spiral symmetry. (c) Point symmetry axis for the arbitrary rotation of the layers $\alpha(z)$.}
    \label{fig:mlstruct}
\end{figure}

We examine a multilayered metamaterial formed by stacking anisotropic layers with the principal components of permittivity tensor $\eps_1$, $\eps_2$ and $\eps_3$. The layers are periodically rotated with respect to each other such that one of their anisotropy axes remains in $Oxy$ plane forming angle $\alpha(z)$ with the $x$ axis, see Fig.~\ref{fig:mlstruct}(a).

If the period of the lattice $a$ is much smaller than the wavelength $\lambda$, the whole structure behaves as an effective medium and its material equations to some approximation can be presented in the form Eq.~\eqref{eq:bianis_form}. The tensors $\hat{\eps}$, $\hat{\mu}$ and $\hat{\kappa}$ entering those equations are constrained by the symmetries of the meta-structure, which we analyze below.

We start from the simplest option when the layers are very thin compared to the lattice period and the anisotropy axis rotates uniformly from layer to layer:
\begin{equation}\label{eq:Spiral}
    \alpha(z) = \frac{\pi z}{a},
\end{equation}
%
%
which we further call spiral symmetry. In such case, the symmetry group of the structure includes $C_2^{(x)}$ rotation with respect to the $x$ axis and the combination of arbitrary rotation $C_{\alpha}^{(z)}$ relative to the $z$ axis followed by the proportional shift [Fig.~\ref{fig:mlstruct}(b)]. Since translations do not affect bulk material parameters, chirality tensor has to satisfy
\begin{equation}\label{eq:transformation}
\hat{\kappa}= \pm\hat{R}(g) \, \hat{\kappa}\,\hat{R}^{-1}(g),
\end{equation}
where $g$ is a symmetry element which is either $C_2^{(x)}$ or $C_{\alpha}^{(z)}$ with arbitrary angle $\alpha$.
$\hat{R}$ is a matrix of the respective transformation and the sign in Eq.~\eqref{eq:transformation} depends on the sign of the determinant $\text{det}\,\hat{R}$, e.g. $+$ for proper rotations, $-$ for reflections. Straightforward calculation yields that in the case of spiral symmetry chirality tensor is necessarily diagonal:
\begin{equation}\label{eq:chirality-spiral}   \hat{\kappa}=\text{diag}\left(\kappa_{11},\kappa_{11},\kappa_{33}\right)\:,
\end{equation}
but not necessarily isotropic, i.e. $\kappa_{11}\not=\kappa_{33}$. In fact, $\kappa_{33}$ component is not manifested in normal incidence geometry and can only be probed at oblique incidence. Note that in practice the combination of symmetries $C_2^{(x)}$ and $C_n^{(z)}$ with $n\geq 3$ suffice to ensure the diagonal structure of chirality tensor, Eq.~\eqref{eq:chirality-spiral}.


In the same spirit, we examine more general scenario with the arbitrary law $\alpha(z)$ of anisotropy axis rotation with the only requirement  $\alpha(0)=\alpha(a)=0$ [Fig.~\ref{fig:mlstruct}(c)]. This lowers the symmetry of the structure down to $C_2^{(z)}$ leaving a single symmetry element and resulting in a more complicated chirality tensor 
\begin{equation}\label{eq:chirality-general}
    \hat{\kappa}^\text{(arb)} = \begin{pmatrix}
        \kappa_{11} & \kappa_{12} & 0\\
        \kappa_{21} & \kappa_{22} & 0\\
        0 & 0 & \kappa_{33}\\
    \end{pmatrix}\:.
\end{equation}
Besides anisotropic chirality, the system develops two more responses: pseudochirality captured by the symmetric zero-trace part of the matrix Eq.~\eqref{eq:chirality-general} and omega-type response characterized by the antisymmetric part of Eq.~\eqref{eq:chirality-general}. The distinction between the two responses also depends on symmetry.

\section{Derivation of the effective material parameters: normal incidence scenario}\label{sec:NormalInc}

In our theoretical analysis, we assume that the first anisotropic layer is located at $z=0$ possessing the permittivity tensor
\begin{equation}
    \label{3_varepsilon_{01}}
\hat{\varepsilon}_{01}=\begin{pmatrix}
        \varepsilon_1 & 0 & 0\\
        0 & \varepsilon_2 & 0\\
        0 & 0 & \eps_3
    \end{pmatrix}.
\end{equation}
Other layers with identical properties are rotated by different angles $\alpha$ in $Oxy$ plane. Hence, the permittivity of the layer with the coordinate $z$ is given by 
\begin{equation}
\begin{split}
        \hat{\varepsilon}(z) & = \hat{R}(\alpha)\hat{\varepsilon}_{01}\hat{R}^{-1}(\alpha)\\
    =&\begin{pmatrix}
       \bar{\eps} & 0 & 0\\
       0 & \bar{\eps} & 0\\
       0 & 0 & \eps_3
       \end{pmatrix} 
    +
    \Delta\varepsilon\begin{pmatrix}
        -\cos(2\alpha) & -\sin(2\alpha) & 0\\
         -\sin(2\alpha) & \cos(2\alpha) & 0\\
         0 & 0 & 0
    \end{pmatrix},\label{3_eps_period}
\end{split}
\end{equation}
where $\hat{R}$ matrix describes counterclockwise rotation with respect to $z$ axis
\begin{equation}\label{eq:rotationmatrix}
\hat{R}(\alpha)=
    \begin{pmatrix}
    \cos\alpha & -\sin\alpha & 0\\
    \sin\alpha & \cos\alpha & 0\\
    0 & 0 & 1
    \end{pmatrix}\:,
\end{equation}
$\bar{\varepsilon}=(\varepsilon_1+\varepsilon_2)/2$ is the average permittivity, $\Delta\varepsilon=(\varepsilon_2-\varepsilon_1)/2$ is the half-difference between the principal permittivity components and $b=2\pi/a$ is the reciprocal lattice period.





In the case of monochromatic fields and monochromatic external sources $\v{j}$ Maxwell's equations yield
\begin{equation}
    \label{3_wave_eq}
    \Delta\v{E}-\nabla(\operatorname{div}\v{E})+q^2\v{D}=-\frac{4\pi\,i\,q}{c}\,\v{j},
\end{equation}
where $q=\omega/c$, $\omega$ is the frequency of the wave, $c$ is the speed of light and CGS system of units is employed.


Periodicity of the system allows to expand the electric field in the form
\begin{equation}
\label{3_solution_of_Max_eq}
    \v{E}(\v{r})=\sum_n \v{E}_n \exp\left[i{\bf k}_\bot\cdot{\bf r}+i(k_z+nb)z\right],
\end{equation}
where $\v{E}_{n}$ are the Floquet harmonics of electric field, $\v{k}$ is the wave vector with in-plane component $\v{k}_\bot$ and the similar expansion is valid for the electric displacement $\v{D}$. 

Deriving the effective medium description, we focus on the behavior of the {\it averaged} fields represented by the zeroth order Floquet harmonic of the expansion Eq.~\eqref{3_solution_of_Max_eq}. Assuming that the external current density $\v{j}$ has only zeroth order Floquet harmonic $\v{j}_0$, we aim to compute higher-order Floquet harmonics in terms of the averaged fields and recover the link between $\v{D}_0$ and $\v{E}_0$, which  defines the effective material parameters of interest. In turn, the presence of external sources $\v{j}$ allows to distinguish frequency and spatial dispersion as $\v{k}$ and $\om$ become independent.

To illustrate the above strategy, we consider first the scenario when the wave vector is orthogonal to the layers, $\v{k}=(0, 0, k_z)^\mathsf{T}$, while electric field lies in the $Oxy$ plane: $\v{E}=(E_x, E_y, 0)^\mathsf{T}$. As the electric field is orthogonal to $\v{k}$ and $\v{k}+nb\,\bas$, the problem simplifies, $\nabla\left(\divv{\v{E}}\right)$ drops out from Eq.~\eqref{3_wave_eq} and we immediately recover the connection between higher-order Floquet harmonics of electric field and displacement:
\begin{equation}
\label{3_Floquet_coeff}
    \v{E}_n=\frac{q^2}{(k_z+nb)^2}\,\v{D}_n \approx \frac{q^2}{n^2 b^2}\left(1-\frac{2k_z}{nb}\right)\v{D}_n+O(\xi^4),
\end{equation}
where $\xi=q/b=a/\lambda$ is a small parameter and $n\neq0$.

On the other hand, electric field and electric displacement at a given point are related to each other via the usual constitutive relation $\v{D}(z)=\hat{\eps}(z)\,\v{E}(z)$. Expanding the permittivity in the Fourier series $\hat{\eps}(z)=\sum\limits_{n}\hat{\eps}_n\,e^{inbz}$ and using the Floquet expansion of the fields, Eq.~\eqref{3_solution_of_Max_eq}, we recover
\begin{equation}\label{3_vector D}
\v{D}_n=\sum_{n'}\hat{\varepsilon}_{n-n'}\v{E}_{n'}=\hat{\varepsilon}_n\v{E}_0+O(\xi^2).
\end{equation}
Combining Eqs.~\eqref{3_Floquet_coeff} and \eqref{3_vector D}, we deduce the averaged displacement in the form

%
\begin{equation*}
    \v{D}_0=\sum_n \hat{\varepsilon}_{-n}\v{E}_n=\hat{\varepsilon}_0\v{E}_0+\sum_{n\neq0}\hat{\varepsilon}_{-n}\v{E}_n\approx
\end{equation*}

\begin{equation*}
    \approx\hat{\varepsilon}_0\v{E}_0+\frac{q^2}{b^2}\sum_{n\neq0}\frac{\hat{\varepsilon}_{-n}\hat{\varepsilon}_n}{n^2}\v{E}_0-\frac{2k_z q^2}{b^3}\sum_{n\neq 0}\frac{\hat{\varepsilon}_{-n}\hat{\varepsilon}_n}{n^3}\v{E}_0=
\end{equation*}

\begin{equation}
\label{3_zero mode of D}
    =\left[\hat{\varepsilon}_0+\frac{q^2}{b^2}{\sum_{n=1}^{\infty}\frac{\{\hat{\varepsilon}_{-n},\hat{\varepsilon}_n\} }{n^2}}\right]\v{E}_0-\frac{2k_z q^2}{b^3}{\sum_{n=1}^{\infty}\frac{[\hat{\varepsilon}_{-n},\hat{\varepsilon}_n]}{n^3}}\v{E}_0,
\end{equation}
where the terms proportional to $\xi^4$ or higher powers of small parameter $\xi$ are neglected.

Such connection between the averaged fields is a typical example of {\it spatial dispersion}, when the electric displacement depends not only on the field at the same point, but also on its spatial derivatives (see the term $\propto k_z$). However, the first derivative of $\v{E}$ yields magnetic field $\v{H}$ and therefore such corrections are equivalent to bianisotropy~\cite{Serdyukov2001} with the  material equations  Eq.~\eqref{eq:bianis_form}. The connection between the nonlocal permittivity tensor $\hat{\epsilon}(\om,\v{k})$ and local effective material parameters $\hat{\eps}$ and $\hat{\kap}$ reads~\cite{Silveirinha2007}:
\begin{equation}\label{eq:Connection}
\hat{\epsilon}(\om,\v{k})=\left[\hat{\eps}-\hat{\kap}\hat{\kap}^\mathsf{T}\right]+\frac{i}{q}\,\left[\hat{\kap}\v{k}^\times+\v{k}^\times\hat{\kap}^\mathsf{T}\right]\:,
\end{equation}
where $\v{k}_{il}=e_{ijl}\,k_j$ and $e_{ijl}$ is a fully antisymmetric tensor with $e_{123}=1$.

For the general structure of chirality tensor Eq.~\eqref{eq:chirality-general} and normal incidence geometry, spatial dispersion correction in Eq.~\eqref{eq:Connection} takes a simple form
\begin{equation}\label{eq:Correction}
    \delta\hat{\epsilon}(\om,\v{k})=2i\kappa_{\text{eff}} k_z/q\,\bas^\times\:,
\end{equation}
where $\kappa_{\text{eff}}=(\kappa_{11}+\kappa_{22})/2$.
%
To proceed further, we evaluate $\hat{\eps}_n$ Fourier components with $n\not=0$ explicitly from Eq.~\eqref{3_eps_period}:
\begin{equation}\label{eq:EpsFourier} \hat{\varepsilon}_{n}\equiv\frac{1}{a} \int_{0}^{a} \hat{\varepsilon}(z) e^{-i n b z} d z = -\Delta\eps\,\left(\sigma_z\,C_n+\sigma_x\,S_n\right)\:,
\end{equation}
%
where we consider $xy$ subspace, $\sigma_x$ and $\sigma_z$ denote Pauli matrices and the dimensionless coefficients $C_n$ and $S_n$ are given by
\begin{gather}
    C_n=\frac{1}{a}\,\int\limits_0^a\,e^{-inbz}\,\cos 2\alpha(z)\,dz\:,\label{eq:Cn}\\
    S_n=\frac{1}{a}\,\int\limits_0^a\,e^{-inbz}\,\sin 2\alpha(z)\,dz\label{eq:Sn}
\end{gather}
and depend on the specific way how the anisotropy axis is rotated from layer to layer.

Straightforward calculation and the comparison of Eqs.~\eqref{3_zero mode of D}, \eqref{eq:Correction} yield the general expression for effective chirality
\begin{equation}\label{eq:ChiralityExpr}
  \kappa_{\text{eff}}=-\frac{4\Delta\eps^2\,q^3}{b^3}\,\sum\limits_{n=1}^{\infty}\,\frac{\Im\left(C_n\,S_n^*\right)}{n^3}\:,  
\end{equation}
%
while $xx$ and $yy$ components of effective permittivity read
\begin{equation}\label{eq:PermittivityExpr}    \eps_{\text{eff}}=\bar{\eps}+\frac{2q^2\Delta\eps^2}{b^2}\,\sum\limits_{n=1}^{\infty}\,\frac{|C_n|^2+|S_n|^2}{n^2}\:.
\end{equation}
%
%

Equations~\eqref{eq:ChiralityExpr},\eqref{eq:PermittivityExpr} define the effective material parameters of a multilayer for the arbitrary law of anisotropy axis rotation. Interestingly, the correction to the effective permittivity appears already in the second order in $\xi=q/b$, while chirality arises only in the third order vanishing in the subwavelength limit $\xi\ll 1$. 

It is instructive to examine the case of spiral medium with a specific law of anisotropy axis rotation, Eq.~\eqref{eq:Spiral}. In this case, for $n\geq 1$ $C_n=1/2\,\delta_{n1}$ and $S_n=-i/2\,\delta_{n1}$ and thus
%

\begin{gather}
    \varepsilon_{\text{eff}} = \bar{\varepsilon}+\Delta\varepsilon^2 \frac{q^2}{b^2},\label{3_inf_large_layers_eps}\\
    \kappa_{\text{eff}}= - \Delta \varepsilon^2\frac{q^3}{b^3}.\label{3_inf_large_layers_kappa}   
\end{gather}

Note that the latter equations were obtained previously~\cite{Vetrov2020,Voronin2024} without using homogenization theory but examining the eigenmodes of a spiral medium instead (Supplementary Materials, Sec.~I). Indeed, electromagnetic chirality is imprinted in the structure of the eigenmodes which have circular polarizations and different refractive indices 
\begin{equation}
    \label{3_two_ref_ind}
    n_{\pm} = \sqrt{\varepsilon \mu} \pm \kappa,
\end{equation}
which allows to probe chirality via eigenmode simulations.

\begin{figure}[b!]
    \centering
    \includegraphics[width=\linewidth]{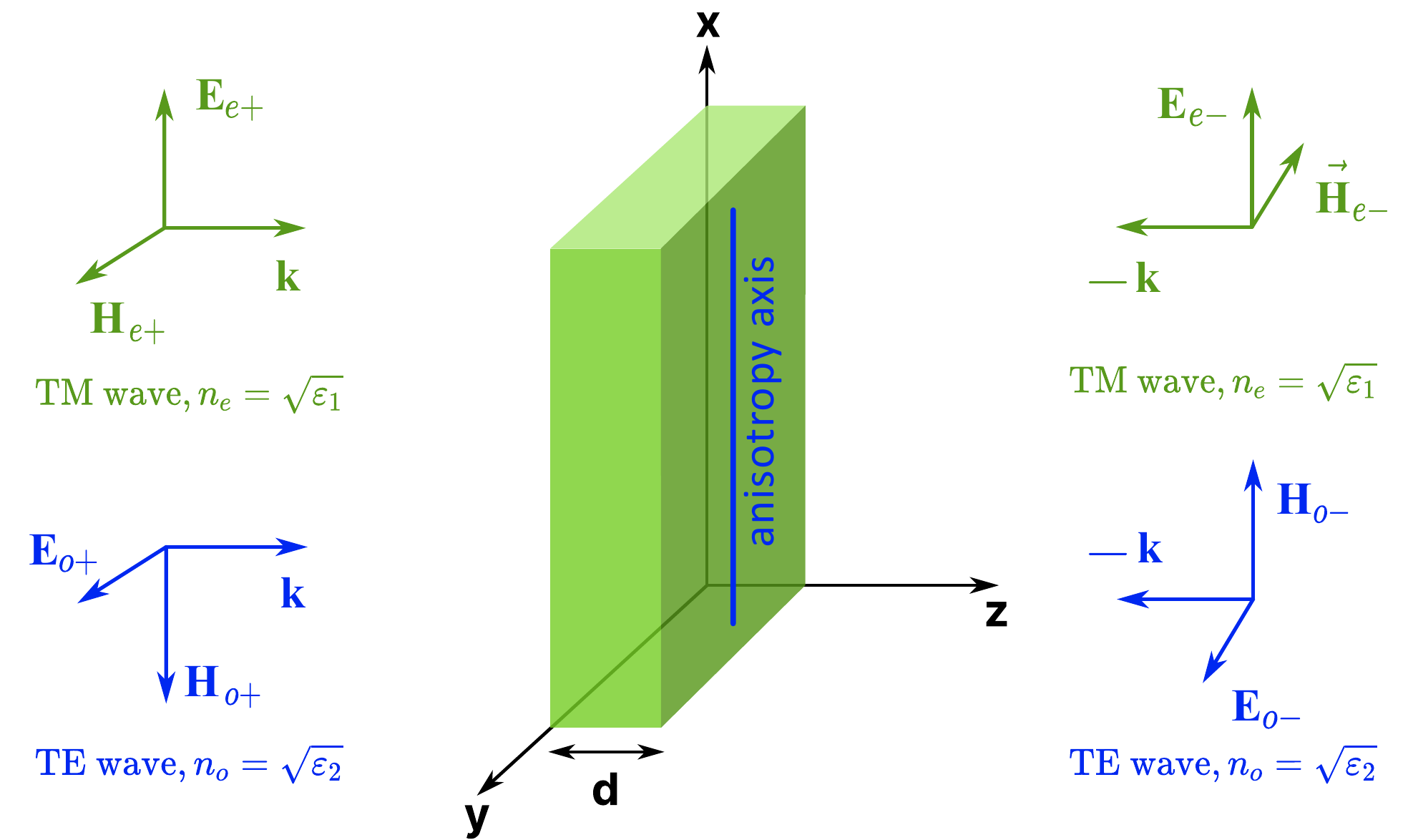}
    \caption{Reflection and transmission of waves in anisotropic dielectric slab and construction of the transfer matrix.}     \label{4fig_transfer_matrxi_method}
\end{figure}

It should be stressed that the solution Eqs.~\eqref{eq:ChiralityExpr}, \eqref{eq:PermittivityExpr} presented here is more general and allows to compute chiral response in other physically relevant situations~-- for instance, when the unit cell of the lattice contains few anisotropic layers and the rotation of anisotropy axis from layer to layer is not continuous but rather stepwise (see Supplementary Materials, Sec.~II).

\section{Validation of the effective medium description}\label{sec:Validation}


To test the validity of the effective medium description, we simulate  multilayered metamaterial using rigorous transfer matrix method. We introduce the vector of the fields $\v{v}=(E_x,E_y,H_x,H_y)^\mathsf{T}$ and define the transfer matrix as 
\begin{equation}\label{4_transf_matrix_basis}
\v{v}(d)=\hat{M}(d)\,\v{v}(0)\:.
\end{equation}
%
%
Then, given the eigenmodes of anisotropic dielectric [Fig.~\ref{4fig_transfer_matrxi_method}], the tranfer matrix $\hat{M}(d)$ for the homogeneous slab with $x$-oriented anisotropy axis reads



\begin{small}
\begin{equation*}
    \label{4_transfer_matrix}
    \begin{pmatrix}
    \cos{(qn_ed)}&0&0&\frac{i}{n_e}\sin{(qn_ed)}\\
        0&\cos{(qn_od)}&-\frac{i}{n_0}\sin{(qn_od)}&0\\
        0&-in_0\sin{(qn_od)}& \cos{(qn_od)}&0\\
in_e\sin{(qn_ed)}&0&0&\cos{(qn_ed)}
    \end{pmatrix}.
\end{equation*}
\end{small}

If the layer is rotated, its transfer matrix is computed as
\begin{equation*}
    \label{4_multipli_matrix}
    \hat{M}_{m} = \hat{T}(\alpha_{m})\hat{M}_{0}\hat{T}^{-1}(\alpha_{m}),
\end{equation*}
with
\begin{equation*}
    \hat{T}(\alpha)=
    \begin{pmatrix}
     R_2(\alpha) & 0_2\\
     0_2 & R_2(\alpha)
    \end{pmatrix}\:,
\end{equation*}
where $R_2(\alpha)$ is the $xy$ block of the rotation matrix Eq.~\eqref{eq:rotationmatrix} and $0_2$ is $2\times 2$ zero matrix.
%
%
The transfer matrix for the lattice period is recovered by multiplying the matrices of all constituent layers
\begin{equation}
    \label{4_full_transfer_matrix}
    \hat{M}_{\text{period}} = \hat{M}_{N}\cdot\hat{M}_{N-1}\cdot ... \cdot \hat{M}_{2}\cdot\hat{M}_{1}\cdot\hat{M}_{0}.
\end{equation}
The eigenmodes of the periodic structure are found from the equation
\begin{equation}\label{eq:eigenmodes}
\left[\hat{M}_{\text{period}}-e^{ik_z a}\right]\,\v{v}(0)=0\:,
\end{equation}
where the eigenvalues $k_z$ are related to the refractive indices of the eigenmodes, while the respective eigenvectors ${\bf v}$ define their polarization at the point $z=0$.

\begin{figure*}[t]
    \centering
    \includegraphics[width=0.9\textwidth]{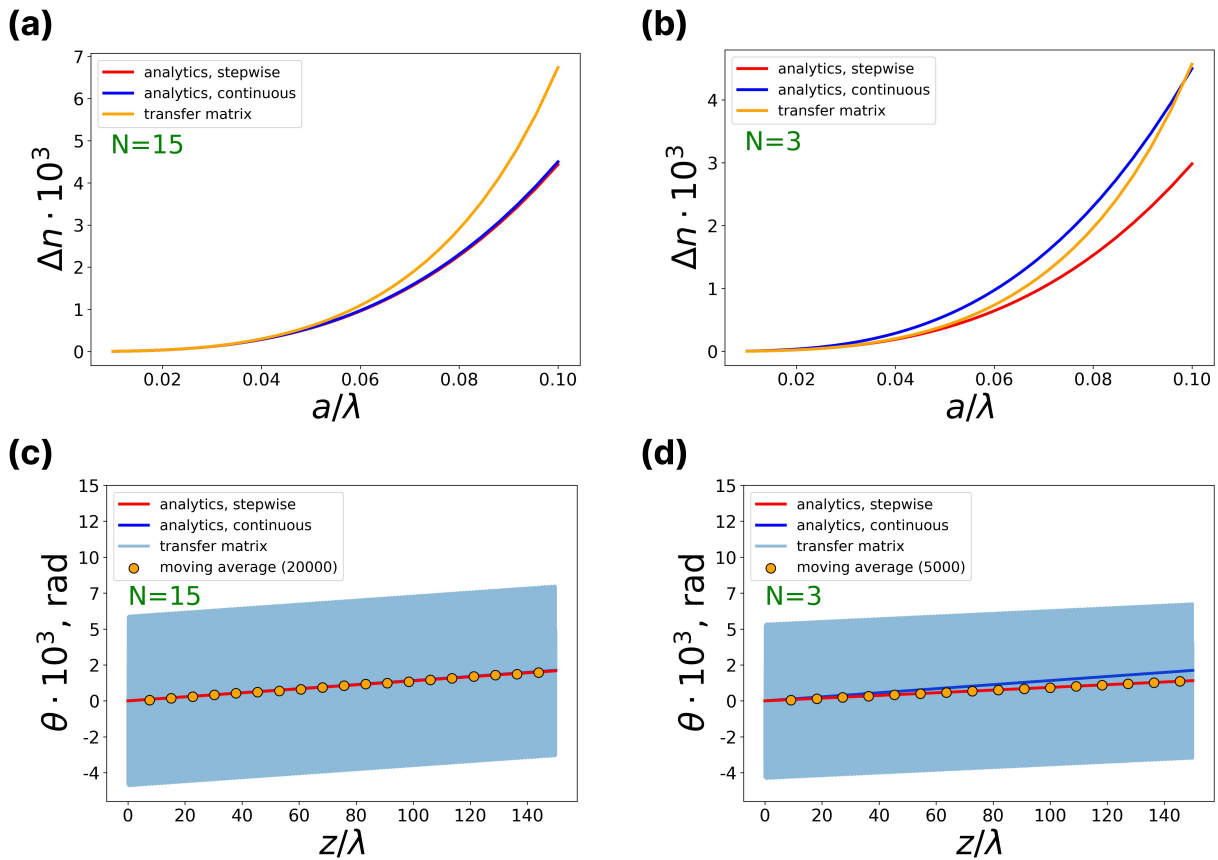}
    \caption{Probing the eigenmodes of a multilayered structure composed of anisotropic layers. (a,b) Refractive index difference for the two forward-propagating modes versus $a/\lambda$ ratio when the period consists of (a) $N=15$ and (b) $N=3$ layers, respectively. Orange, blue and red line depict the results of rigorous transfer matrix simulation and analytical calculation for the continuous and stepwise anisotropy axis rotation, respectively. (c,d) Polarization plane rotation for the linearly polarized wave propagating in the bulk metamaterial with $a/\lambda = 0.01$. Light blue line depicts the results of transfer matrix simulation, orange dots correspond to the moving average over 200 and 500 layers, 10 dots are taken within each layer. Blue and red curves show the analytical results for the continuous and stepwise rotation of anisotropy axis. The permittivity of the layers is $\varepsilon_1 = 10$, $\varepsilon_2 = 7$. The overall number of periods is equal to 15000.
    }
    \label{fig:validation_normal}
\end{figure*}


We solve Eq.~\eqref{eq:eigenmodes} for experimentally relevant parameters~\cite{Voronin2024} at normal incidence ($k_z=nq$). Assuming that the lattice period is formed by $N=15$ anisotropic layers with a uniform rotation of anistropy axis from layer to layer, in Fig.~\ref{fig:validation_normal}(a) we plot the difference of the refractive indices of the two forward-propagating eigenmodes which quantifies the effective chirality [see Eq.~\eqref{3_two_ref_ind}]. As chirality is proportional to $(a/\lambda)^3$, $\Delta n$ is tiny 
in the deeply subwavelength region $\xi=a/\lambda\ll 1$. However, with the increase of $\xi$ up to 0.06, $\Delta n$ grows rapidly. At this point, the developed perturbation theory is no longer valid, and the discrepancy between the effective medium model and transfer matrix method mounts. At the same time, the assumption about the continuous rotation of anisotropy axis remains good approximation for $N=15$ layers in the period.

A qualitatively similar behavior of $\Delta n$ is observed, when the period of the lattice includes just $N=3$ layers [Fig.~~\ref{fig:validation_normal}(b)]. In this case, however, one has to take into account that the rotation of anisotropy axis from layer to layer is not continuous, but rather stepwise, as further analyzed in the Supplementary Materials, Sec.~II. 

Another aspect of chirality is the rotation of polarization plane of the linearly polarized wave propagating in the chiral medium. Decomposing the wave into two circularly polarized modes and taking into account the difference of the refractive indices, Eq.~\eqref{3_two_ref_ind}, we recover the rotation $\theta=-q\kappa_{\text{eff}}\,z$. The results for $N=15$ and $N=3$ layers in the period for the fixed ratio $a/\lambda=0.01$ are presented in Fig.~\ref{fig:validation_normal}(c,d). We compare these results with the transfer matrix calculation. In the latter case, however, polarization varies strongly within the layer, being elliptical in the general case. We compute the angle of rotation at each point as
\begin{equation*}
    \label{eq: numeric rotation}
    \theta = \frac{1}{2}\arctan{\left(\frac{2 E_{tx} E_{ty} \cos(\Delta\phi)}{E_{tx}^2 - E_{ty}^2}\right)},
\end{equation*}
where $E_{tx}$ and $E_{ty}$ are the amplitudes of light transmitted through each layer and $\Delta\phi$ is the phase difference between those components. These results are depicted by the thin blue line in Fig.~\ref{fig:validation_normal}(c,d) creating a uniform shading due to the rapid oscillations of polarization within the metamaterial period. Next we average the rotation angle taking a sliding averaging window. The obtained averaged results are in full agreement with the analytical theory demonstrating the rotation of polarization plane linearly growing with distance indicating the onset of effective chirality.




\section{Maximization of effective chirality}\label{sec:Maximization}

To illustrate the power of the obtained general solution for chirality $\kappa_{\text{eff}}$, we explore the following problem. Assume that the multilayered structure is composed of very thin layers with the given principal components of permittivity tensor $\eps_1$, $\eps_2$, and the period $a$ of the lattice is fixed. In this setting, emergent chirality depends on the orientation of the layers described by the periodic function $\alpha(z)$. Below, we aim to find the arrangement of the layers maximizing chiral response $\kappa_{\text{eff}}$.

To do so, we compare Eqs.~\eqref{3_zero mode of D} and \eqref{eq:Correction} and present chirality in the form
\begin{equation}\label{eq:ChiralityGen}
    \kappa_{\text{eff}} \, \v{e}_z^\times = i \frac{q^3}{b^3}\sum_{n\not=0} \frac{1}{n^3}\hat{\varepsilon}_{-n} \hat{\varepsilon}_{n}\:.
\end{equation}

For our analysis, it is convenient to transform Eq.~\eqref{eq:ChiralityGen} replacing the summation of 
the Fourier harmonics by the integration in the real space. Assuming that arbitrary periodic function $f(z)$ is presented in the form
\begin{equation*}
f(z) = \sum_{n=-\infty}^{\infty} f_n e^{i n b z}    
\end{equation*}
we introduce the operator $\mathcal{J}$ acting in the Fourier space as
\begin{equation}\label{Ioperator}
 (\mathcal{J} f)_n = \begin{cases}
        \frac{1}{ib n} f_n, &n \neq 0 \\
        0, &n=0.
    \end{cases}   
\end{equation}
Essentially, this substracts the average from the function $f(z)$, integrates it over the interval $[0,z]$ and substracts the average from the resulting function. Accordingly, in the real space the operator $\mathcal{J}$ acts as
\begin{gather}
\mathcal{J} f(z)=\int\limits_{0}^a\,K_1\left(\frac{z-z'}{a}\right)\,f(z')\,dz'\:,\label{eq:Convolution}\\
K_1(\xi)=\frac{1}{2}\,\text{sign}\left(\xi\right)-\xi\:.
\end{gather}
Having Eq.~\eqref{eq:Convolution}, it is straightforward to compute the repeated action of the operator $\mathcal{J}$. For instance,
\begin{gather}
\mathcal{J}^3 f(z)=a^2\,\int\limits_{0}^a\,K_3\left(\frac{z-z'}{a}\right)\,f(z')\,dz'\:,\label{eq:Convolution3}\\
K_3(\xi)=-\frac{1}{12}\,\xi\left(1-3|\xi|+2\xi^2\right)\:.
\end{gather}

Yet another ingredient needed for our calculation are the definitions of the scalar product and norm
\begin{align}
    \langle f, g \rangle &= \int_0^a f^*(z) g(z) dz = a \sum_{n = -\infty}^\infty f_n^* g_n, \\
    \|f\|^2 &= \langle f,f \rangle = a \sum_{n = -\infty}^\infty |f_n|^2\:.
\end{align}
As the permittivity tensor of the layers is real, $\hat{\eps}_{-n}=\hat{\eps}_n^*$ and therefore
\begin{gather}
\kappa_{\text{eff}} \, \v{e}_z^\times = q^3\,\sum_{n\not=0} \hat{\varepsilon}_{n}^*\,\frac{\hat{\eps}_n}{(ibn)^3}=q^3\,\sum_{n\not=0} \hat{\eps}_n^*\,\left(\mathcal{J}^3\hat{\eps}\right)_n\notag\\
=\frac{q^3}{a}\int\limits_0^a\hat{\eps}(z)\,\left[\mathcal{J}^3\hat{\eps}(z)\right]\,dz\notag\\
=q^3\,a\int\limits_0^a\int\limits_0^a\,\hat{\eps}(z)\,K_3\left(\frac{z-z'}{a}\right)\,\hat{\eps}(z')\,dz'dz\notag\\
=\frac{q^3\,a}{2}\int\limits_0^a\int\limits_0^a\,K_3\left(\frac{z-z'}{a}\right)\,\left[\hat{\eps}(z),\hat{\eps}(z')\right]\,dz'dz\:. 
\end{gather}

Straightforward calculation of the commutator using Eq.~\eqref{3_eps_period} yields
\begin{equation}
 [\hat{\varepsilon}(z), \hat{\varepsilon}(z')] = 2 \Delta \varepsilon^2 \Im\left(e^{2i \alpha(z)}\,e^{-2i\alpha(z')}\right) \v{e}_z^\times   
\end{equation}
giving an explicit formula for chirality
\begin{gather}
\kappa_{\text{eff}}=q^3 a\Delta\eps^2\,\Im\int\limits_0^a\int\limits_0^a\,K_3\left(\frac{z-z'}{a}\right)\,e^{2i\alpha(z)}\,e^{-2i\alpha(z')}\,dz dz'\notag\\
=\frac{q^3}{a}\Delta\eps^2\,\Im\left<\psi,\mathcal{J}^3\psi\right>\:,
\end{gather}
where $\psi(z)=\exp(-2i\alpha(z))$ is an auxiliary scalar function. The latter expression can be readily estimated as follows:
\begin{align*}
    |\Im \langle \psi, \mathcal{J}^3 \psi \rangle| 
    & = a \left|\Im \sum_{n \neq 0} \frac{\psi^*_n \psi_n}{-i b^3 n^3}\right| \\
    & \leq \frac{a}{b^3} \sum_{n \neq 0} \left|\frac{1}{n^3}\right| \psi^*_n \psi_n\leq \frac{a}{b^3} \sum_{n} \psi^*_n \psi_n\\
    & = \frac{1}{b^3} \|\psi\|^2=\frac{a}{b^3}
\end{align*}

Hence, an upper bound for chirality \(\kappa_{\text{eff}}\) reads:
\begin{equation} \label{eq:chirality maximum}
    |\kappa_\text{eff}| \leq \frac{q^3}{b^3} \Delta \varepsilon^2\:. 
\end{equation}
This value is achieved for $\psi=e^{-ibz}$. Thus, we conclude that the arrangement of the layers maximizing chiral response is the continuous rotation of anisotropy axis, i.e. spiral symmetry.

\begin{figure}
    \centering
    \includegraphics[width=\linewidth]{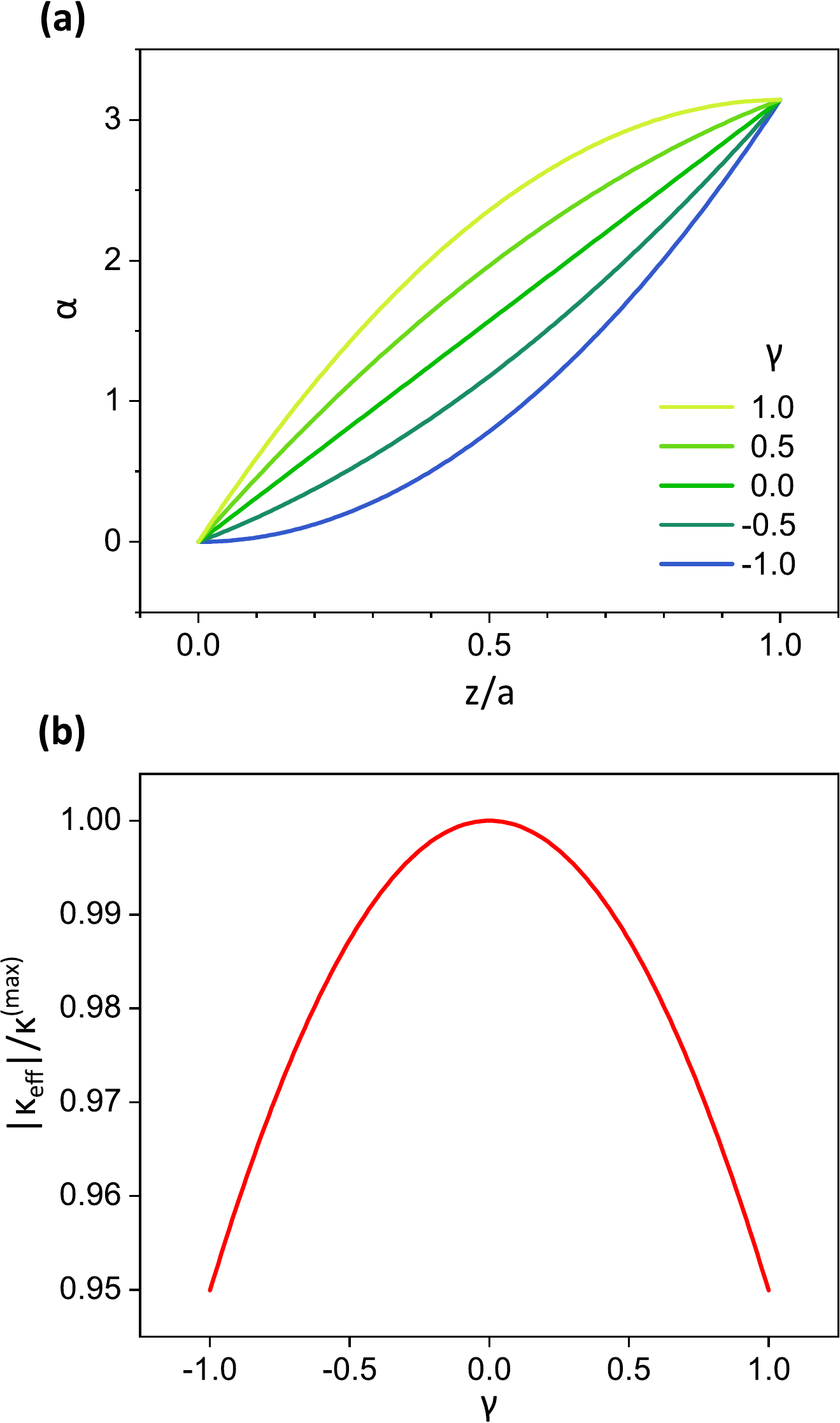}
    \caption{Illustration of effective chirality maximization. (a) Various laws of anisotropy axis rotation \(\alpha(z)\) distinguished by the $\gamma$ parameter. (b) Relative effective chirality as a function of \(\gamma\). Chirality reaches its maximum for the uniform rotation of anisotropy axis, i.e. in case of spiral symmetry.}
    \label{fig:chirality vs gamma}
\end{figure}

To illustrate that, we chose a specific set of  \(\alpha(z)\) functions, parameterized by \(\gamma\) as follows:
\begin{equation}
    \alpha_\gamma(z) = \pi \left[ \frac{z}{a} + \gamma \frac{z}{a} \left(1 - \frac{z}{a} \right) \right]
\end{equation}
In Fig. \ref{fig:chirality vs gamma} we plot the dependence of effective chirality \(\kappa_{\text{eff}}\) on \(\gamma\). We chose a range \([-1,1]\) for \(\gamma\), so that the rotation angle $\alpha(z)$ changes monotonously with $z$. As anticipated from our general analysis, chiral response reaches its maximum exactly for \(\gamma=0\).

\section{Reconstruction of the full chirality tensor}\label{sec:ObliqueInc}

In this section, we generalize our treatment and examine an arbitrary direction of $\v{k}$ vector. In experimental situation, this corresponds to the oblique incidence scenario. For simplicity, we assume that the anisotropy axis rotates continuously according to Eq.~\eqref{eq:Spiral}, so that the effective chirality tensor has the diagonal structure, Eq.~\eqref{eq:chirality-spiral}. While $\kappa_{11}$ component is calculated above, $\kappa_{33}$ term is only manifested at oblique incidence.

As we consider an unbounded medium with spiral symmetry, without loss of generality we can rotate the coordinate system and assume that the wave vector is $\v{k}=(k_x, 0, k_z)^\mathsf{T}$, while all components of electric field are nonzero $\v{E}=(E_x, E_y, E_z)^\mathsf{T}$. In such setting, the wave equation Eq.~\eqref{3_wave_eq} results in the system: 
\begin{gather}
-(k_z+nb)^2 E_{nx}+k_x(k_z+nb)E_{nz}+q^2 D_{nx}=0,\label{6_mixed_1}\\
k_x(k_z+nb)E_{nx}-k_x^2 E_{nz}+q^2 D_{nz}=0,\label{6_mixed_2}\\
-(k_z+nb)^2 E_{ny}-k_x^2 E_{ny}+q^2 D_{ny}=0.\label{6_mixed_3}
\end{gather}
%
%
Equations \eqref{6_mixed_1}
and \eqref{6_mixed_2} can be combined to yield
\begin{equation}
    \label{6_div D}
    k_x D_{nx}+(k_z+nb) D_{nz}=0.
\end{equation}

In turn, the material equation for the Floquet harmonics of electric field reads
\begin{equation}
    \label{6_floquet coefficients}
     \v{D}_{n}=\sum_{m} \hat{\varepsilon}_{n-m}
 \v{E}_{m},
\end{equation}
or, alternatively, introducing the inverse permittivity $\hat{\zeta}$,
\begin{equation}
    \label{6_floquet coefficients2}
     \v{E}_{n}=\sum_{m} \hat{\zeta}_{n-m}
 \v{D}_{m}.
\end{equation}



Next we employ the perturbation theory with  the small parameter $\xi=a/\lambda$ presenting the Floquet harmonics with $n\not=0$ as a series
\begin{equation}
    \label{6_peturbation theory}
    \begin{cases}
        E_{ni}=E_{ni}^{(0)}+\xi E_{ni}^{(1)}+\xi^2 E_{ni}^{(2)}+...\\
        D_{ni}=D_{ni}^{(0)}+\xi D_{ni}^{(1)}+\xi^2 D_{ni}^{(2)}+...
    \end{cases}
\end{equation}
where index $i=x,y,z$ labels the Cartesian component.

Performing the calculations similarly to the normal incidence case (see details in Supplementary Materials, Section~III), we recover the effective material parameters:
\begin{equation*}
    \label{6_final_eff_params_eps}
    \hat{\varepsilon}_{\text{eff}} = \begin{pmatrix}
        \bar{\varepsilon}+\frac{\Delta \varepsilon^{2}}{2 b^{2}}\left(2 q^{2}-\frac{k_{x}^{2}}{\varepsilon_{3}}\right)&0&0\\
        0&\bar{\varepsilon}+\frac{\Delta \varepsilon^{2}}{2 b^{2}}\left(2 q^{2}-\frac{k_{x}^{2}}{\varepsilon_{3}}\right)&0\\
        0&0&\varepsilon_3
    \end{pmatrix},
\end{equation*}

\begin{equation}
    \label{6_final_eff_params_kappa}
    \hat{\kappa}_{\text{eff}} = -\frac{q}{2} \frac{\Delta\varepsilon^2}{b^3}\left(2q^2 - \frac{k_x^2}{\varepsilon_3}\right)\begin{pmatrix}
        1&0&0\\
        0&1&0\\
        0&0&-1
    \end{pmatrix}.
\end{equation}

Interestingly, effective permittivity tensor $\hat{\eps}_{\text{eff}}$ acquires second-order spatial dispersion correction, while effective chirality is cubic in $\xi$ and also contains spatial dispersion contribution. Equation~\eqref{6_final_eff_params_kappa} suggests that $\kappa_{33}$ component is nonzero and has different sign compared to $\kappa_{11}$.

To validate such behavior of effective chiral response, we analyze both refractive indices and polarizations of the modes supported by the structure at oblique incidence.

\begin{figure}
    \centering
    \includegraphics[width=0.9\linewidth]{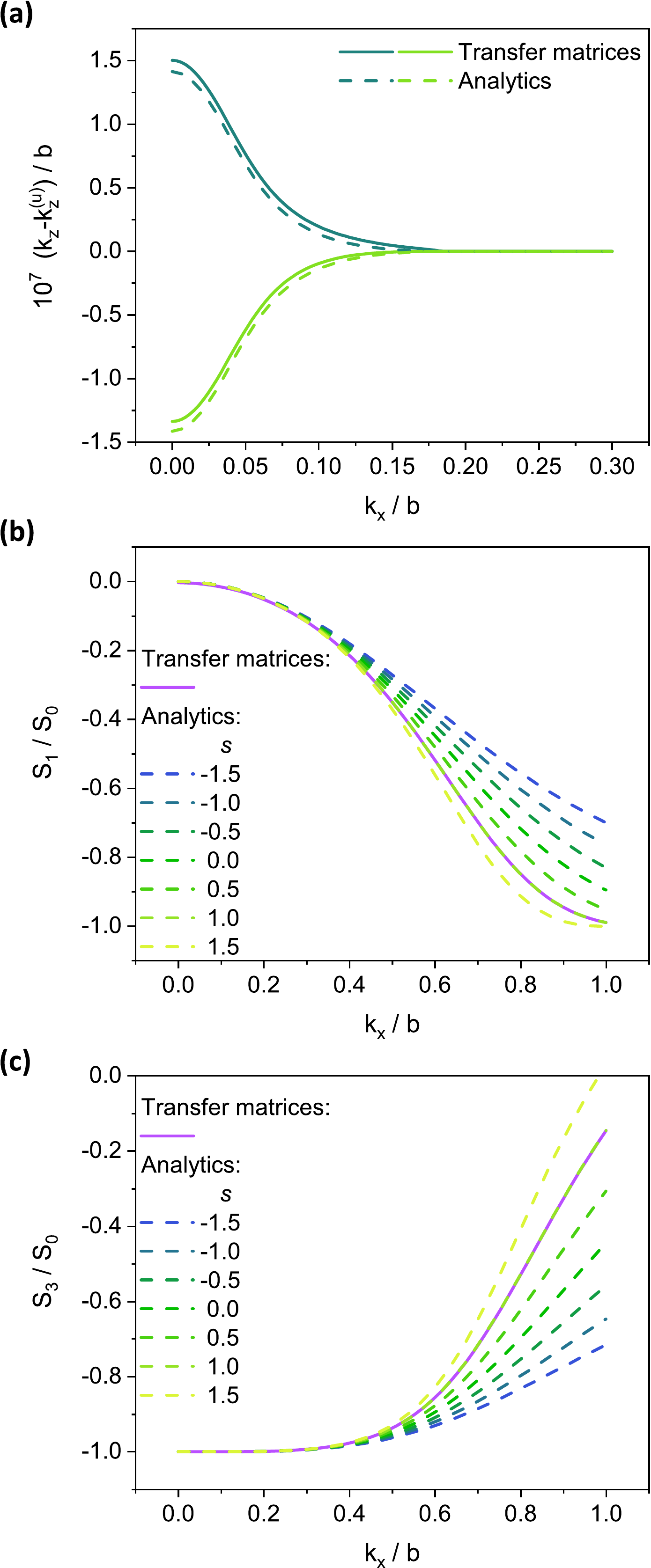}
    \caption{Validation of effective medium description at oblique incidence. 
    (a) Isofrequency contour in $Oxz$ plane with $k_z^{(u)}$ for a uniaxial crystal subtracted. $a/\lambda=0.01$ is fixed, the permittivity values for a single layer are \(\varepsilon_1 = 10.0\), \(\varepsilon_2 = 7.0\), \(\varepsilon_3 = 8.5\). (b, c) Stokes parameters for a single eigenmode with the larger \(k_z\), calculated by the transfer matrix method, superposed with analytical results corresponding to different values of \(\kappa_{33}\). $a/\lambda=0.06$, layer permittivities are the same as above.}
    \label{fig:kappa zz verification}
\end{figure}

First, we fix the frequency $q=\om/c$ and calculate the isofrequency contour in the $Oxz$ plane. To isolate the effects due to chirality and spatial dispersion, we subtract from the propagation constant $k_z$ its dominant part \(k_z^\text{(u)}\)~-- a value obtained for a uniaxial non-chiral crystal with
\begin{equation}
    \hat{\varepsilon}^\text{(u)} = \text{diag}\left(\bar{\varepsilon},\bar{\varepsilon},\varepsilon_3\right), \quad \hat{\kappa}^\text{(u)} = \hat{0}\:.
\end{equation}
Figure~\ref{fig:kappa zz verification}(a) shows the corrections to the propagation constants of two forward-propagating (\(k_z > 0\)) modes at different incidence angles quantified by the $k_x$. We observe that the effective medium model correctly reproduces $k_z$ values at all incidence angles.

To support our effective medium model further, we calculate the polarizations of the same modes quantifying them by the Stokes parameters defined as
%
%
\begin{align}
    S_0 &= |E_{0x'}|^2 + |E_{0y'}|^2, \\
    S_1 &= |E_{0x'}|^2 - |E_{0y'}|^2, \\
    S_2 &= 2 \Re \left( E_{0x'} E_{0y'}^*\right), \\
    S_3 &= - 2 \Im \left( E_{0x'} E_{0y'}^*\right).
\end{align}
Here, $\v{E}_0$ denotes the averaged field calculated as
\begin{equation*}
    \v{E}_0 = \frac{1}{a}\int_0^a \v{E}(z) e^{- i k_z z} dz,
\end{equation*}
and \(k_z\) is recovered from the eigenmode calculation,  Eq.\eqref{eq:eigenmodes}. The Stokes parameters are evaluated in the plane $Ox'y'$ spanned by \(\Re \v{E}_0\) and \(\Im \v{E}_0\) vectors; \(Ox'\) corresponds to the projection of \(Ox\) axis onto the polarization plane, while $Oy'$ axis satisfies the conditions
\[
    [\v{e}_{x'} \times \v{e}_{y'}] \cdot \v{e}_{z} > 0, \quad \v{e}_{x'} \cdot \v{e}_{y'} = 0.
\]
This choice of the coordinate system ensures that \(S_2 = 0\).


In our calculation, we choose \(\varepsilon_3 = \bar{\varepsilon}\) in order to mitigate the effect of optical anisotropy in the limit \(\xi \to 0\) and isolate the influence of \(\kappa_{33}\) chirality component. We perform the effective medium calculation for various \(k_x\) and for the set of chirality tensors with different values of $\kappa_{33}$ component: $\kappa_{33}=-s\,\kappa_{11}$ comparing the results for a mode with a larger $k_z$ in Fig.~\ref{fig:kappa zz verification}(b,c). Comparing the analytical results with the transfer matrix calculation, we recover that the best approximation corresponds to $s=1$, meaning that $\kappa_{33}=-\kappa_{11}$ in agreement with the theoretical prediction.

%
%

\section{Discussion and conclusions}\label{sec:Discussion}

In summary, we have derived a complete effective medium description of the multilayered structure composed of anisotropic layers rotated relative to each other. Our description captures the metamaterial regime when the lattice period is much smaller than the wavelength. In such case, electromagnetic chirality is proportional to $\left(a/\lambda\right)^3$ and features spatial dispersion contribution at oblique incidence.

As we proved, continuous rotation of anisotropy axis from layer to layer maximizes the effective chirality compared to any other configuration with the same lattice period $a$ and permittivities of the individual layers $\eps_1$ and $\eps_2$.

While our analysis provides insights into the emergence of effective chirality in a stack of anisotropic layers, we believe that our methodology is more general and enables understanding of other layered geometries featuring more exotic responses. It could also pave a way to the reverse engineering of metamaterial structures with the prescribed effective properties.

\section*{Acknowledgments}

We acknowledge Daniel Bobylev for valuable discussions. Theoretical models were supported by the Russian Science Foundation (grant No.~23-72-10026). Numerical investigation of the structure was supported by the Russian Science Foundation (grant No.~24-72-10069). Analysis of chiral response maximization was supported by Priority 2030 Federal Academic Leadership Program. M.A.G. acknowledges partial support from the Foundation for the Advancement of Theoretical Physics and Mathematics ``Basis''.

\bibliography{ChiralLib}

\end{document}


\title{Supplemental Materials: \\ Effective chiral response of anisotropic multilayered metamaterials}

\author{Konstantin Baranov}
\email{These authors have contributed equally to this work}
\affiliation{School of Physics and Engineering, ITMO University, Saint  Petersburg 197101, Russia}

\author{Dmitry Vagin}
\email{These authors have contributed equally to this work}
\affiliation{School of Physics and Engineering, ITMO University, Saint  Petersburg 197101, Russia}

\author{Maxim A. Gorlach}
\email{m.gorlach@metalab.ifmo.ru}
\affiliation{School of Physics and Engineering, ITMO University, Saint  Petersburg 197101, Russia}

\maketitle
\widetext
\tableofcontents

\section{Polarization and refractive indices of the eigenmodes\\ in a helicoidal medium}


In this section, we revisit the problem of light propagation in a dielectric medium with spiral symmetry, i.e. the medium invariant under the combination of infinitesimal rotation and the proportional translation. We treat this problem analogously to Refs.~\cite{Vetrov2020,Voronin2024} generalizing the approach to the arbitrary local material of the layers and computing the refractive indices of the eigenmodes and their polarizations.


We start with Maxwell's equations in a monochromatic case and in the absence of sources written in the CGS system of units
\begin{equation}
    \begin{cases}
    \nabla \times \v{E} = i q \v{B},  \\ 
    \nabla \times \v{H} = -i q \v{D},  \\
    \nabla \cdot \v{D} = 0, \\
    \nabla \cdot \v{B} = 0, \\
    \end{cases}
\end{equation}
where $q=\om/c$. If the fields change strictly along $Oz$ axis, i.e. normal incidence scenario is realized, then
\begin{equation}\label{eq:zpropag}
    \begin{cases}
    \v{e}_z \times \partial_z \v{E} = i q \v{B},  \\ 
    \v{e}_z \times \partial_z \v{H} = -i q \v{D},  \\
    \v{e}_z \cdot \partial_z  \v{D} = 0, \\
    \v{e}_z \cdot \partial_z  \v{B} = 0. \\      
    \end{cases}
\end{equation}
The full system of equations governing the evolution of the fields is composed of Maxwell's equations Eq.~\eqref{eq:zpropag} and the system of constitutive relations, generally written as
\begin{equation}
    \begin{cases}
    \v{D} = \hat{\epsilon} \v{E} + \hat{\alpha} \v{H} \\
    \v{B} = \hat{\beta} \v{E} + \hat{\mu} \v{H},    
    \end{cases}
\end{equation}
where $\hat{\eps}$, $\hat{\mu}$, $\hat{\alpha}$ and $\hat{\beta}$ denote local material parameters. Next we introduce a 6-component vector of electric and magnetic fields constructing it from the components continuous at the boundary between the different media:
\begin{equation}
    \v{F} = (E_x, E_y, D_z, H_x, H_y, B_z)^T,
\end{equation}
and obtain the first-order differential equation
\begin{equation}\label{eq:maineq}
    \partial_z  \v{F} = -i\hat{\mathcal{T}} \v{F}
\end{equation}
with matrix \(\hat{\mathcal{T}}\) expressed via the coefficients of \(\hat{\epsilon}\), \(\hat{\mu}\), \(\hat{\alpha}\), \(\hat{\beta}\). While general expression for \(\hat{\mathcal{T}}\) is cumbersome, the result for anisotropic layer with the permittivity tensor $\hat{\eps}=\text{diag}\left(\eps_1,\eps_2,\eps_3\right)$ is readily recovered from Eq.~\eqref{eq:zpropag} and reads
\begin{equation}
    \hat{\mathcal{T}}_\text{loc} = q \left(
        \begin{array}{cccccc}
         0 & 0 & 0 & 0 & -1 & 0 \\
         0 & 0 & 0 & 1 & 0 & 0 \\
         0 & 0 & 0 & 0 & 0 & 0 \\
         0 & \eps_2 & 0 & 0 & 0 & 0 \\
         -\eps_1 & 0 & 0 & 0 & 0 & 0 \\
         0 & 0 & 0 & 0 & 0 & 0 \\
        \end{array}
        \right)
\end{equation}
Physically, \(\hat{\mathcal{T}}\) matrix provides a generator of translations along the $z$ axis and embeds all the relevant properties of the medium under study. Accordingly, the values of $\v{F}$ at two close points are related as
\begin{equation*}
    \v{F}(z + dz)=\left(1 - i \; dz \; \mathcal{T}(z)\right)\,\v{F}(z)\:.
\end{equation*}
By construction, the object \(F\) transforms as a direct sum of two vectors under rotations \(\hat{R}(\theta)\) around \(z\) axis:
\begin{equation*}
 F' = \begin{pmatrix}
        \hat{R}(\theta) & 0 \\
        0 & \hat{R}(\theta) \\ 
    \end{pmatrix} F = \hat{R}^{(2)}(\theta) \; F,   
\end{equation*}
where the rotation matrix is given by
\begin{equation}\label{eq:srotationmatrix}
\hat{R}(\alpha)=
    \begin{pmatrix}
    \cos\theta & -\sin\theta & 0\\
    \sin\theta & \cos\theta & 0\\
    0 & 0 & 1
    \end{pmatrix}\:.
\end{equation}
In a similar manner to $\hat{\mathcal{T}}$, we introduce \(\hat{\mathcal{R}}\)~-- a generator of infinitesimal \(\hat{R}^{(2)}\) rotation:
\begin{equation}
 \hat{R}^{(2)}(\theta + d\theta) = (\hat{I} + i \; d\theta \;\hat{\mathcal{R}} ) \hat{R}^{(2)}(\theta)\:.
\end{equation}
Explicit calculation yields
\begin{equation}
 \hat{\mathcal{R}}=\begin{pmatrix}
     0 & i & 0 & 0 & 0 & 0\\
     -i & 0 & 0 & 0 & 0 & 0\\
     0 & 0 & 0 & 0 & 0 & 0\\
     0 & 0 & 0 & 0 & i & 0\\
     0 & 0 & 0 & -i & 0 & 0\\
     0 & 0 & 0 & 0 & 0 & 0
 \end{pmatrix}\:.
\end{equation}
As $\hat{\mathcal{R}}$ is independent of $\theta$, the rotation for the arbitrary angle $\theta$ can be written as
\begin{equation}
    \hat{R}^{(2)}(\theta)=\exp\left(i\theta\hat{\mathcal{R}}\right)\:.
\end{equation}
Now we utilize the symmetry of the structure. Spiral symmetry of the medium implies that its properties are invariant under $dz$ translation accompanied by  $d\theta=\pi dz/a=b dz/2$ rotation. Technically, this means that the generator $\hat{\mathcal{T}}(z)$ capturing the properties of the medium depends on $z$ as follows:
\begin{equation}\label{eq:translationgen}
     \quad \hat{\mathcal{T}}(z) = \hat{R}^{(2)}(bz/2) \hat{\mathcal{T}}_{\text{loc}} \hat{R}^{(2)}(-bz/2) \:,
\end{equation}
where $\hat{\mathcal{T}}_{\text{loc}}=\hat{\mathcal{T}}(0)$. Our goal now is to solve the differential equation Eq.~\eqref{eq:maineq} with the $z$-dependent translation generator $\mathcal{T}$ given by Eq.~\eqref{eq:translationgen}. To do this, we consider the fields in a local coordinate system, i.e. the coordinate system associated with the principal axes of the respective layer:
\begin{equation}
    \v{F}_\text{loc} = \hat{R}^{(2)}(-bz/2) \v{F}\:.
\end{equation}
Next, we derive the differential equation for $\v{F}_{\text{loc}}$:
\begin{equation}
\begin{aligned}
    \partial_z \v{F}_\text{loc} &= \left( \partial_z \hat{R}^{(2)}(-bz/2) \right) \v{F} + \hat{R}^{(2)}(-bz/2) \; \partial_z \v{F} \\
    &= - i\frac{b}{2}\hat{\mathcal{R}} \;  \hat{R}^{(2)}(-bz/2) \; \v{F} - \hat{R}^{(2)}(-bz/2) \; i\hat{\mathcal{T}}(z) \; \v{F} \\
    &= - i\frac{b}{2}\hat{\mathcal{R}} \;  \hat{R}^{(2)}(-bz/2) \; \v{F} -  i\hat{\mathcal{T}}_\text{loc} \;\hat{R}^{(2)}(-bz/2)  \; \v{F} 
\end{aligned}
\end{equation}
Thus, we recover that the fields in a local coordinate frame satisfy the differential equation with constant coefficients:
\begin{equation}\label{eq:local}
    \partial_z \v{F}_\text{loc} = -i\left(\hat{\mathcal{T}}_\text{loc}+\frac{b}{2}\hat{\mathcal{R}}\right) \v{F}_\text{loc}
\end{equation}
The solution to this differential equation is a matrix exponent
\begin{equation} \label{eq:spiral symm general solution}
    \v{F}_\text{loc}(z) = \exp \left\{-iz \left( \hat{\mathcal{T}}_\text{loc}+\frac{b}{2} \hat{\mathcal{R}}\right) \right\} F(0), \quad \v{F}(z) = \exp\left\{iz \frac{b}{2} \hat{\mathcal{R}}\right\} \exp \left\{-iz \left(\hat{\mathcal{T}}_\text{loc}+\frac{b}{2} \hat{\mathcal{R}}\right) \right\} \v{F}(0)\:.
\end{equation}

Equation~\eqref{eq:spiral symm general solution} now allows to compute the eigenmodes of the medium. Periodicity of the metamaterial guarantees that there is a basis of solutions satisfying $\v{F}(z+a)=\v{F}(z)\,e^{ika}$, where $k$ is the Bloch wave number related to the effective refractive index as $k=qn$. These Bloch modes are the eigenvectors of \(\hat{\mathcal{T}}_\text{loc}+\frac{b}{2} \hat{\mathcal{R}}\). Indeed, if \(\Tilde{\v{F}}_j\) satisfies
\[
    \left(\hat{\mathcal{T}}_\text{loc}+\frac{b}{2} \hat{\mathcal{R}}\right) \Tilde{\v{F}}_j = \lambda_j \Tilde{\v{F}}_j
\]
then by substituting \(\v{F}(0) = \Tilde{\v{F}}_j\) into Eq.~\eqref{eq:spiral symm general solution} we obtain
\begin{equation} 
    \v{F}(a) = \hat{R}^{(2)}(\pi) \;  e^{-i \lambda_j a} \; \v{F}(0)
\end{equation}
and recover the refractive index of every mode up to a multiple of \(2 \pi / (qa) \):
\begin{equation} \label{eq:spiral refr index ambig}
    q n_j = -\lambda_j + \frac{\pi - 2 \pi l}{a},
\end{equation}
where $l$ is an arbitrary integer. The limit at \(\xi \to 0\), where \(\xi = q/b\), allows us to resolve the ambiguity. In this limit \(q n_j \to 0\) and therefore
\begin{equation} \label{eq:spiral refr index}
    q n_j(\xi) = -\lambda_j(\xi) + \lim_{\xi' \to 0} \lambda_j (\xi').
\end{equation}
We note that \(\lim_{\xi' \to 0} \lambda_j (\xi')\) are eigenvalues of \(\frac{b}{2} \hat{\mathcal{R}}\), meaning those can be represented as multiples of \(\pi / a \), showing consistency between Eq.~\eqref{eq:spiral refr index ambig} and Eq.~\eqref{eq:spiral refr index}.  
By applying Eq.~\eqref{eq:spiral refr index} to the stack of anisotropic dielectiric layers, described in the main text of this Article, we find two refractive indices for forward-propagating ($k>0$) modes:
\begin{equation} \label{eq:spir refr index anis}
	n_- = \frac{b-\sqrt{-4 q \sqrt{b^2 \bar{\epsilon}+\Delta \epsilon ^2 q^2}+b^2+4 q^2 \bar{\epsilon}}}{2 q}, \quad n_+ = \frac{\sqrt{4 q \sqrt{b^2 \bar{\epsilon}+\Delta \epsilon ^2 q^2}+b^2+4 q^2 \bar{\epsilon}}-b}{2 q}\:,
\end{equation}
where $\bar{\eps}=(\eps_1+\eps_2)/2$ and $\Delta\eps=(\eps_2-\eps_1)/2$ as in the main text. A simple series expansion yields
\[
	n_\pm = \sqrt{\epsilon _0} + \frac{\Delta \epsilon ^2 q^2}{2 b^2 \sqrt{\epsilon _0}} \mp \frac{\Delta \epsilon ^2 q^3}{b^3} + O\left(\xi^4 \right).
\]
When compared with the dispersion relation in chiral medium
\[
	n_\pm = \sqrt{\eps_\text{eff}} \pm \kappa_\text{eff},
\]
this allows us to evaluate the effective material parameters:
\begin{equation} \label{spiral eff parameters}
	\eps_\text{eff} = \bar{\epsilon} + \frac{\Delta \epsilon ^2 q^2}{b^2} + O\left(\xi^3 \right), \quad \kappa_\text{eff} = - \frac{\Delta \epsilon ^2 q^3}{b^3} + O\left(\xi^4 \right).
\end{equation}
These values precisely match those obtained in the main text via rigorous homogenization approach. To consistently determine the sign of chirality in Eq.~\eqref{spiral eff parameters}, we consider the averaged field of each eigenmode, calculating
\begin{equation}
    \v{F}_0 = \frac{1}{a} \int_0^a \v{F}(z) e^{-i n q z} \; dz.
\end{equation}
This is easily accomplished by decomposing \(\hat{\mathcal{R}}\) into a linear combination of projection matrices and rewriting Eq.~\eqref{eq:spiral symm general solution} with \(\v{F}(0) = \Tilde{\v{F}}_j\):
\[
    \hat{\mathcal{R}} = \sum_l \mu_l \hat{P}_l, \quad \hat{P}_l^2 = \hat{P}_l, \quad \sum_l \hat{P}_l = \hat{I}.
\]
\[
    \v{F}(z) = \sum_l \exp\left\{i z \frac{b}{2} \mu_l\right\} \hat{P}_l \exp\left\{-i z \lambda_j \right\} \Tilde{\v{F}}_j
\]
\begin{equation} \label{eq:spir homog mode}
    \v{F}_0^{(j)} = \sum_l \left( \frac{1}{a}\int_0^a \exp\left\{i z \frac{b}{2} \mu_l - i z \lim_{\xi->0} \lambda_j(\xi) \right\}  \right) \hat{P}_l \; \Tilde{\v{F}}_j
\end{equation}

In the latter equation the value of refractive index is substituted from Eq.~\eqref{eq:spiral refr index}. Only one summand in Eq.~\eqref{eq:spir homog mode} remains for each \(j\). We denote its index by \(l_j\), so that 
\[
    i \frac{b}{2} \mu_{l_j} - i \lim_{\xi->0} \lambda_j(\xi) = 0.
\] 
The final expression:
\begin{equation} \label{eq:spir homog simpl}
    \v{F}_0^{(j)} = \hat{P}_{l_j} \; \Tilde{\v{F}}_j
\end{equation}
Projection matrices \(\hat{P}_{l_j}\) filter out either one of two circularly polarized waves or a longitudinal wave. We identify that in  Eq.~\eqref{eq:spir refr index anis} \(n_+\) corresponds to the right-handed circularly polarized mode and \(n_-\) corresponds to the left-handed one. While local fields \(\v{F}(z)\) are elliptically polarized, homogenized modes are polarized circularly. This result is general, the equation~\eqref{eq:spir homog simpl} does not depend on local constitutive relations, but rather follows from the symmetry of the arrangement.

The procedure described above is applicable to the case of oblique incedence as well. However simple analytical solutions, arising due to the constant coefficients in Eq.~\eqref{eq:local}, are no longer available. The local translation generator in this case reads
\[
    \hat{\mathcal{T}}_\text{loc} = \left(
        \begin{array}{cccccc}
         0 & 0 & -\frac{k_{\text{loc}, x}}{\varepsilon _3} & 0 & -q & 0 \\
         0 & 0 & -\frac{k_{\text{loc}, y}}{\varepsilon _3} & q & 0 & 0 \\
         k_{\text{loc}, x} \; \varepsilon _1 & k_{\text{loc}, y} \; \varepsilon _2 & 0 & 0 & 0 & 0 \\
         0 & q \varepsilon _2 & 0 & 0 & 0 & -k_{\text{loc}, x} \\
         -q \varepsilon _1 & 0 & 0 & 0 & 0 & -k_{\text{loc}, y} \\
         0 & 0 & 0 & k_{\text{loc}, x} & k_{\text{loc}, y} & 0 \\
        \end{array}
        \right)
\]
where $\v{k}_{\text{loc}}$ is defined as
\[
    \v{k}_{\text{loc}} = \hat{R}(-bz/2)\; \v{k}.
\]

\section{Derivation of non-zero Fourier components of permittivity tensor for the case of normal incidence and its analysis}

\subsection{Derivation of non-zero Fourier components of the dielectric permittivity tensor}

 Fourier components  can be written in the following form
\begin{equation} 
\label{S2_two integrals}
\hat{\varepsilon}_{n}=\frac{1}{a} \int_{0}^{a} \hat{\varepsilon}(z) e^{-i n b z} d z = \frac{1}{a}\left[\int_{0}^{a}\bar{\varepsilon}\hat{I}e^{-i n b z} d z + \Delta\varepsilon  \int_{0}^{a}\left(\begin{array}{cc}
-\cos (2 \alpha) & -\sin (2 \alpha) \\
-\sin (2 \alpha) & \cos (2 \alpha)
\end{array}\right)e^{-i n b z} d z\right].
\end{equation}

First integral in the right hand side of \eqref{S2_two integrals} vanishes, if $n\neq0$
\begin{equation}
    \label{S2_first integral}
    \int_{0}^{a}\bar{\varepsilon}\hat{I}e^{-i n b z} d z = \begin{cases}
        \bar{\varepsilon} \hat{I}a, \quad n=0 \\
        0, \qquad n \neq 0 
    \end{cases}.
\end{equation}

Since the angle $\alpha$, describing the rotation of the anisotropy axis, has period $\pi$, for the $m$-th layer it is determined as 
\begin{equation}
    \alpha_m = \frac{\pi(m-1)}{N},
\end{equation}

\noindent where $N$ is the number of layers in the period, $m \geq 1$. 

\begin{figure}[h!]
    \centering
    \includegraphics[scale=0.25]{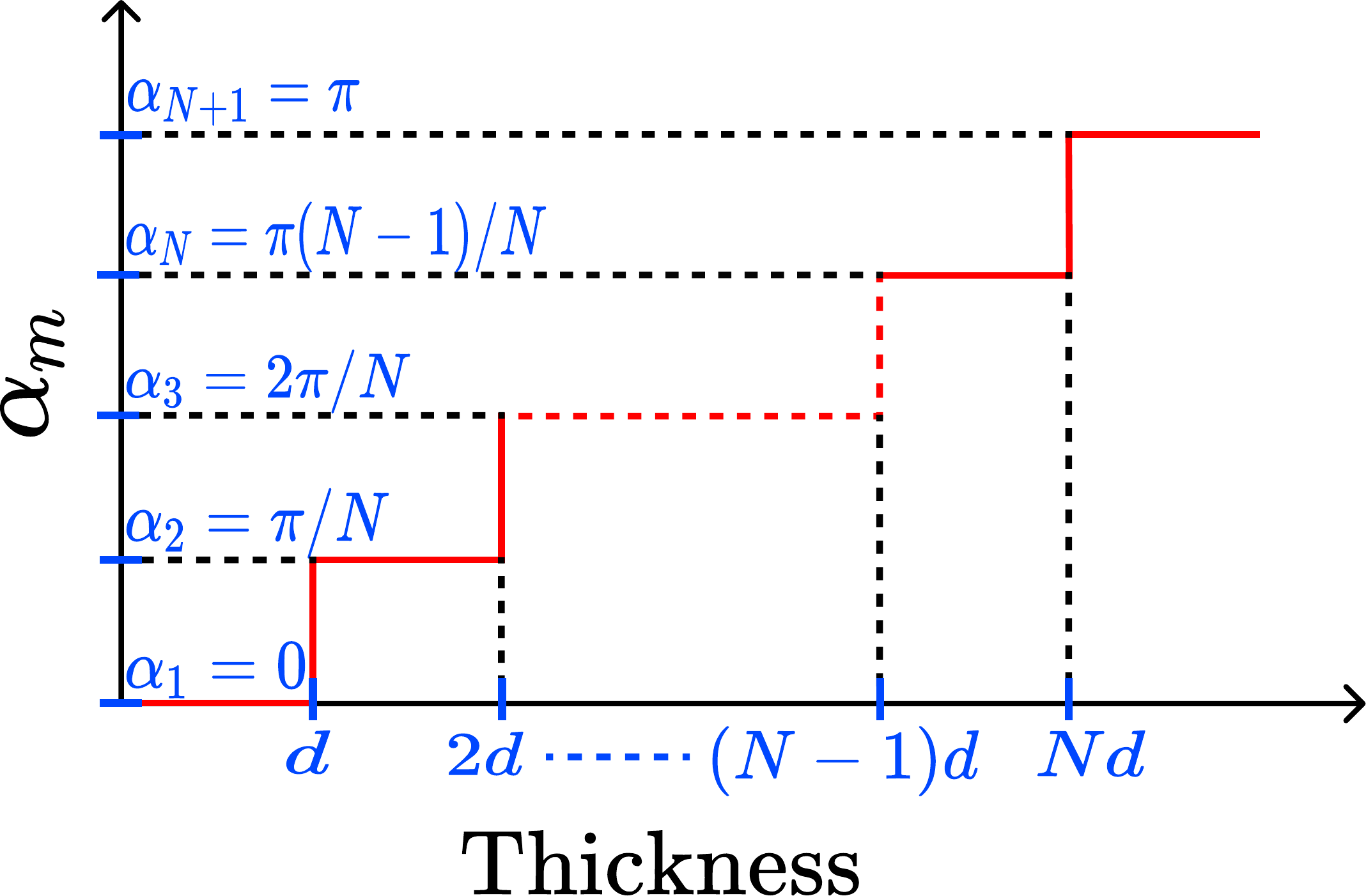}
    \caption{The illustration of the stepwise anisotropy axis rotation}
    \label{step_rot}
\end{figure}

Fig. \ref{step_rot} shows the dependence of anisotropy axis rotation from layer to layer on the thickness of multilayer structure. 

Now the second integral from equation \eqref{S2_two integrals}, containing the angle $\alpha$ in the argument of the trigonometric functions, can be divided into $N$ integrals:

\renewcommand{\arraystretch}{1.75} 

\begin{equation*}
    \int_{0}^{a}\left(\begin{array}{cc}
-\cos (2 \alpha) & -\sin (2 \alpha) \\
-\sin (2 \alpha) & \cos (2 \alpha)
\end{array}\right)e^{-i n b z} d z = 
\int_{0}^{d}\left(\begin{array}{cc}
-\cos (0) & -\sin (0) \\
-\sin (0) & \cos (0)
\end{array}\right)e^{-i n b z} d z + 
\end{equation*}

\renewcommand{\arraystretch}{1.75} 

\begin{equation}
\label{S2_sec_int_1}
+\int_{d}^{2d}\left(\begin{array}{cc}
-\cos (2\pi/N) & -\sin (2\pi/N) \\
-\sin (2\pi/N) & \cos (2\pi/N)
\end{array}\right)e^{-i n b z} d z + ... + \int_{(N - 1)d}^{Nd = a}\left(\begin{array}{cc}
-\cos \left(2 \frac{\pi(N-1)}{N}\right) & -\sin \left(2 \frac{\pi(N-1)}{N}\right) \\
-\sin \left(2 \frac{\pi(N-1)}{N}\right) & \cos \left(2 \frac{\pi(N-1)}{N}\right)
\end{array}\right)e^{-i n b z} d z,
\end{equation}

\noindent where $d = a/N$ is a thickness of layers. If $n=0$, each integral from equation \eqref{S2_sec_int_1} reduces to $d$.

Therefore, taking into account \eqref{S2_first integral} and \eqref{S2_sec_int_1} for $n=0$, we obtain the following expression

\renewcommand{\arraystretch}{1.75} 

\begin{equation}
\label{S2_eps_0_M}
    \hat{\varepsilon}_0 = \bar{\varepsilon}\hat{I} + \Delta \varepsilon \frac{d}{a} \left(\begin{array}{cc}
M_{11} & M_{12} \\
M_{21} & M_{22}
\end{array}\right).
\end{equation}

\renewcommand{\arraystretch}{1.75}  


Next we calculate the elements $M_{11}$, $M_{12}$, $M_{21}$ and $M_{22}$  from \eqref{S2_eps_0_M} individually

\begin{equation}
\label{S2_M11_1}
    M_{22} = - M_{11} = \left[1 + \cos(2\pi/N) + \cos(4\pi/N)  + ... + \cos(2(N-1)\pi/N)\right] =  \sum_{n=0}^{N-1} \cos(2\pi n/N) =  \delta_{1, N},
\end{equation}

\begin{equation}
\label{S2_M12_M_21_1}
    M_{12} = M_{21} = - \left[0 + \sin(2\pi/N) + \sin(4\pi/N)  + ... + \sin(2(N-1)\pi/N)\right] = -\sum_{n=0}^{N-1} \sin(2\pi n/N) = 0.
\end{equation}

The combination of \eqref{S2_eps_0_M}-\eqref{S2_M12_M_21_1} yields

\renewcommand{\arraystretch}{1} 

\begin{equation}
     \hat{\varepsilon}_0 = \bar{\varepsilon}\hat{I} + \Delta \varepsilon 
     \left(\begin{array}{cc}
-1 & 0 \\
0 & 1
\end{array}\right)\delta_{1, N}.
\end{equation}

Now consider the case when $n\neq 0$. Due to Eq.~\eqref{S2_first integral}, only the integral \eqref{S2_sec_int_1} needs to be calculated. Examine a few integrals to establish the dependence

\begin{equation}
    \int_{0}^{d} e^{-inbz} dz = e^{-i\pi nd/a}\frac{a}{\pi n} \sin\left(\frac{\pi n d}{a}\right),
\end{equation}

\begin{equation}
    \int_{d}^{2d} e^{-inbz} dz = e^{-3i\pi nd/a}\frac{a}{\pi n} \sin\left(\frac{\pi n d}{a}\right),
\end{equation}

\begin{equation}
    \int_{2d}^{3d} e^{-inbz} dz = e^{-5i\pi nd/a}\frac{a}{\pi n} \sin\left(\frac{\pi n d}{a}\right),
\end{equation}

\begin{equation}
    \int_{(N-1)d}^{Nd} e^{-inbz} dz = e^{-(2N-1)i\pi n d / a}\frac{a}{\pi n} \sin\left(\frac{\pi n d}{a}\right).
\end{equation}

After substituting these values into equation \eqref{S2_sec_int_1}, we obtain a complex expression

\renewcommand{\arraystretch}{1.75} 

\begin{equation*}
    \hat{\varepsilon}_n = \frac{\Delta \varepsilon}{\pi n} \sin\left(\frac{\pi n d}{a}\right) \left(\begin{array}{cc}
L_{11} & L_{12} \\
L_{21} & L_{22}
\end{array}\right).
\end{equation*}





Similar to expressions \eqref{S2_M11_1} and \eqref{S2_M12_M_21_1}, the elements $L_{11}$, $L_{12}$, $L_{21}$ and $L_{22}$ can be calculated separately

\begin{equation*}
    L_{22} = -L_{11} = e^{-i\pi n d / a} + \cos \left(2\frac{\pi}{N}\right)e^{-3i\pi n d / a} + \cos \left(2\frac{2\pi}{N}\right)e^{-5i\pi n d / a} + ... + \cos \left(2\frac{(N-1)\pi}{N}\right)e^{-(2n-1)i\pi n d / a} = 
\end{equation*}

\begin{equation*}
    = \sum_{m=0}^{N-1}\cos\left(2\frac{\pi m}{N}\right)e^{-i(2m+1)\pi n / N}= \sum_{m=0}^{N-1}\frac{1}{2}e^{-i\pi n / N}\left(e^{i[2m\pi (1-n) / N]} +  e^{-i[2m\pi (1+n) / N]}\right) = 
\end{equation*}

\begin{equation}
   =\frac{N}{2}e^{-i\pi n / N}\left(\sum_{k\in \mathbb{Z}}\delta_{1-Nk, n} + \sum_{k\in \mathbb{Z}}\delta_{Nk-1, n}\right),  
\end{equation}

\begin{equation*}
    L_{12} = L_{21} = - \left(\sin \left(2\frac{\pi}{N}\right)e^{-3i\pi n d / a} + \sin \left(2\frac{2\pi}{N}\right)e^{-5i\pi n d / a} + ... + \sin \left(2\frac{(N-1)\pi}{N}\right)e^{-(2n-1)i\pi n d / a}\right)=
\end{equation*}

\begin{equation}
    = -\sum_{m=1}^{N-1}\sin\left(2\frac{\pi m}{N}\right)e^{-i(2m+1)\pi n / N} = i\frac{N}{2}e^{-i\pi n / N}\left(\sum_{k\in \mathbb{Z}}\delta_{1-Nk, n} - \sum_{k\in \mathbb{Z}}\delta_{Nk-1, n}\right).
\end{equation}

\noindent where $k$ is an integer number.

Therefore, in general case the Fourier components for $n\neq0$ can be presented as 


\begin{equation}
\label{S2_n neq0_final}
    \hat{\varepsilon}_{n} = \frac{N}{2}\frac{\Delta\varepsilon}{\pi n}\sin\left(\frac{\pi n}{N}\right)e^{-i\pi n / N}\left[\sum_{k\in \mathbb{Z}} \delta_{1-Nk, n} \begin{pmatrix}
        -1 & i \\ i & 1
    \end{pmatrix} + \sum_{k\in \mathbb{Z}} \delta_{Nk-1, n} \begin{pmatrix}
        -1 & -i \\ -i & 1
    \end{pmatrix} \right].
\end{equation}

If \(N > 2\), then for every \(n\) no more than one term within two sums is nonzero. We can write
\begin{equation}
    \hat{\varepsilon}_{1-Nk} = -(-1)^k \frac{N}{2} \frac{\Delta\varepsilon}{\pi (Nk-1)} \sin\left(\frac{\pi}{N}\right) \begin{pmatrix}
        -1 & i \\ i & 1
    \end{pmatrix}, \quad \hat{\varepsilon}_{N k - 1} = (-1)^k \frac{N}{2}\frac{\Delta\varepsilon}{\pi (N k - 1)}\sin\left(\frac{\pi}{N}\right) \begin{pmatrix}
        -1 & -i \\ -i & 1
    \end{pmatrix}
\end{equation}
for any \(k \in \mathbb{Z}\) and if \(n\) cannot be presented as \(1 - N k\) or \(Nk - 1 \), then the permittivity component \(\hat{\varepsilon}_n\) vanishes.




\subsection{Calculation of sums with commutative and anticommutative relations}

The main text of the paper presents the results of computing the infinite sums, commutators and anticommutators involving Fourier components described by Eq.~\eqref{S2_n neq0_final}. The commutation and anticommutation relations can be represented in the following form

\begin{equation}
    [\hat{\varepsilon}_{Nk - 1}, \hat{\varepsilon}_{1-Nk}] =-i\frac{N^2 \Delta\varepsilon^2}{\pi^2(Nk-1)^2} \sin^2(\frac{\pi}{N})\begin{pmatrix}
        0 & -1 \\ 1 & 0
    \end{pmatrix},
\end{equation}
\begin{equation}
    [\hat{\varepsilon}_{1 - Nk}, \hat{\varepsilon}_{N k - 1}] = i\frac{N^2 \Delta\varepsilon^2}{\pi^2(Nk-1)^2} \sin^2(\frac{\pi}{N})\begin{pmatrix}
        0 & -1 \\ 1 & 0
    \end{pmatrix},
\end{equation}
\begin{equation}
    \{\hat{\varepsilon}_{N k - 1}, \hat{\varepsilon}_{1-Nk}\} = -\frac{N^2 \Delta\varepsilon^2}{\pi^2(Nk-1)^2} \sin^2(\frac{\pi}{N})\begin{pmatrix}
        1 & 0 \\ 0 & 1
    \end{pmatrix},
\end{equation}
\begin{equation}
    \{{\hat{\varepsilon}_{1-Nk}, \hat{\varepsilon}_{N k - 1}\}}= -\frac{N^2 \Delta\varepsilon^2}{\pi^2(Nk-1)^2} \sin^2(\frac{\pi}{N})\begin{pmatrix}
        1 & 0 \\ 0 & 1
    \end{pmatrix}
\end{equation}


\noindent and performed the sum with commutator


\begin{equation*}
\sum_{n=1}^{\infty}\frac{[\hat{\varepsilon}_{-n},\hat{\varepsilon}_n]}{n^3}=\sum_{1-Nk >0}^{\infty} \frac{[\hat{\varepsilon}_{Nk - 1}, \hat{\varepsilon}_{1-Nk}]}{(1-Nk)^3} + \sum_{Nk-1>0}^{\infty}\frac{[\hat{\varepsilon}_{1 - Nk}, \hat{\varepsilon}_{N k - 1}]}{{(Nk-1)^3}} = 
\end{equation*}

\begin{equation*}
    =-i\frac{N^2 \Delta\varepsilon^2}{\pi^2}\sin^2(\frac{\pi}{N})\begin{pmatrix}
        0 & -1 \\ 1 & 0
    \end{pmatrix}\left[\sum_{1-Nk >0}^{\infty}\frac{1}{(1-Nk)^5} - \sum_{Nk - 1 > 0}^{\infty}\frac{1}{(Nk-1)^5}\right]
\end{equation*}

\begin{equation*}
    =-i\frac{N^2 \Delta\varepsilon^2}{\pi^2}\sin^2(\frac{\pi}{N})\begin{pmatrix}
        0 & -1 \\ 1 & 0
    \end{pmatrix}\left[\sum_{k = 0}^{\infty}\frac{1}{(1+Nk)^5} - \sum_{k=1}^{\infty}\frac{1}{(Nk-1)^5}\right] = 
\end{equation*}

\begin{equation}
    =-
 i\frac{\Delta\varepsilon^2 N^2}{\pi^2}\begin{pmatrix}
    0&-1\\
    1&0
\end{pmatrix}\frac{\sin^2\left(\frac{\pi}{N}\right)}{24 N^5}\left[\psi^{(4)}\left(\frac{1}{N}\right) - \psi^{(4)}\left(1-\frac{1}{N}\right)\right]
\end{equation}

\noindent and, analogously, the sum with anticommutator

\begin{equation}
\sum_{n=1}^{\infty}\frac{\{\hat{\varepsilon}_{-n},\hat{\varepsilon}_n\}}{n^2} =  \frac{\Delta\varepsilon^2 N^2}{\pi^2}\begin{pmatrix}
        1&0\\
        0&1  \end{pmatrix}\frac{\sin^2\left(\frac{\pi}{N}\right)}{6 N^4}\left\{\psi^{(3)}\left(\frac{1}{N}\right) + \psi^{(3)}\left(1-\frac{1}{N}\right)\right\},
\end{equation}

\noindent where $\psi^{(m)}(z)$ is the polygamma function of order $m$, which is defined through the gamma function as follows

\begin{equation*}
    \psi^{(m)}(z) = \frac{d^{m+1}}{dz^{m+1}}\operatorname{ln}\Gamma(z).
\end{equation*}

As a result, the effective parameters for stepwise  rotation of anisotropy axis can be written in the following form

\begin{equation}
\label{S2_effect_mat_param_2}
    \hat{\varepsilon}_{\text{eff}} = \operatorname{\hat{I}}\bar{\varepsilon}+\Delta\varepsilon\begin{pmatrix}
-1&0\\
0&1
\end{pmatrix}\delta_{1, N}+\operatorname{\hat{I}}\frac{q^2}{b^2}\frac{\Delta\varepsilon^2}{\pi^2}\frac{\sin^2\left(\frac{\pi}{N}\right)}{6N^2}\left\{\psi^{(3)}\left(\frac{1}{N}\right) + \psi^{(3)}\left(1-\frac{1}{N}\right)\right\},
\end{equation}

\begin{equation}
    \label{S2_effect_mat_param}
    \hat{\kappa}_{\text{eff}} =  \operatorname{\hat{I}}\frac{q^3}{b^3}\frac{\Delta\varepsilon^2 }{\pi^2}\frac{\sin^2 \left(\frac{\pi}{N}\right)}{24N^3}\left[\psi^{(4)}\left(\frac{1}{N}\right) - \psi^{(4)}\left(1-\frac{1}{N}\right)\right].
\end{equation}

Analysing the limiting case ($N \longrightarrow \infty$) in equations \eqref{S2_effect_mat_param_2} and \eqref{S2_effect_mat_param}, we recover the material effective parameters for the medium with the continuous and uniform rotation of the anisotropy axis

\begin{equation}
\label{S2_inf_large_layers_eps}
    \varepsilon_{\text{eff}}' = \bar{\varepsilon}+\Delta\varepsilon^2 \frac{q^2}{b^2},
\end{equation}
\begin{equation}
\label{S2_inf_large_layers_kappa}   
    \kappa_{\text{eff}}'= - \Delta \varepsilon^2\frac{q^3}{b^3}.
\end{equation}

\section{Derivation of the full chirality tensor}

In this section, we derive full chirality tensor examining a general scenario with the arbitrary direction of the wave vector ${\bf k}$. Combining Floquet expansion with the wave equation, we recover
\begin{equation}
    \label{S3_system of eq}
    \begin{cases}
        -(k_z+nb)^2 E_{nx}+k_x(k_z+nb)E_{nz}+q^2 D_{nx}=0,\\
        k_x(k_z+nb)E_{nx}-k_x^2 E_{nz}+q^2 D_{nz}=0,\\
        -(k_z+nb)^2 E_{ny}-k_x^2 E_{ny}+q^2 D_{ny}=0.
    \end{cases}
\end{equation}

If the anisotropy axis rotates with coordinate as $\alpha(z)=\pi z/a$, then the Fourier components of permittivity tensor read 
\begin{equation}
    \label{S3_non-zero varep comp}
  \hat{\varepsilon}_0=\begin{pmatrix}
      \bar{\varepsilon} & 0 & 0\\
      0 & \bar{\varepsilon} & 0\\
      0 & 0 & \varepsilon_3
  \end{pmatrix},  \qquad \hat{\varepsilon}_1=\frac{\Delta \varepsilon}{2}\begin{pmatrix}
    -1 & i & 0\\
    i & 1 & 0\\
    0 & 0 & 0
\end{pmatrix},                                  \qquad\hat{\varepsilon}_{-1}=\frac{\Delta \varepsilon}{2}\begin{pmatrix}
    -1 & -i & 0\\
    -i & 1 & 0\\
    0 & 0 & 0
\end{pmatrix}.
\end{equation}

\noindent At the same time, the Fourier components of inverse permittivity tensor take the form

\begin{equation}
    \label{S3_non-zero xi comp}
  \hat{\zeta}_0=\frac{1}{\varepsilon_1 \varepsilon_2}\begin{pmatrix}
      \bar{\varepsilon} & 0 & 0\\
      0 & \bar{\varepsilon} & 0\\
      0 & 0 & \frac{\varepsilon_1 \varepsilon_2}{\varepsilon_3}
  \end{pmatrix},  \qquad \hat{\zeta}_1=\frac{\Delta \varepsilon}{2\varepsilon_1 \varepsilon_2}\begin{pmatrix}
    1 & -i & 0\\
    -i & -1 & 0\\
    0 & 0 & 0
\end{pmatrix},                                  \qquad\hat{\zeta}_{-1}=\frac{\Delta \varepsilon}{2\varepsilon_1 \varepsilon_2}\begin{pmatrix}
    1 & i & 0\\
    i & -1 & 0\\
    0 & 0 & 0
\end{pmatrix}.
\end{equation}


Introducing expansion parameter $\xi$ explicitly, we recover the system of equations

\begin{equation}
    \label{S3_system_floquet}
    \begin{cases}
        (\xi k_z+nb)^2E_{nx}-\xi k_x(\xi k_z+nb)E_{nz}=\xi^2q^2D_{nx}, \vspace{0.2 cm}\\
       E_{n z}= \left(\zeta_{n}\right)_{z i} D_{0 i}+\sum_{m \neq 0}  \left(\zeta_{n-m}\right)_{z i} D_{m i},\vspace{0.2 cm}\\
        D_{n x}= \left(\varepsilon_{n}\right)_{x i} E_{0 i}+\sum_{m \neq 0}  \left(\varepsilon_{n-m}\right)_{x i} E_{m i},\vspace{0.2 cm}\\
        \xi k_x D_{nx}+(\xi k_z+nb) D_{nz}=0.
    \end{cases}
\end{equation}

Expanding the Floquet harmonics of the fields in powers of small parameter, we obtain the following system of equations 

\begin{equation}
    \label{S3_peturbation theory}
    \begin{cases}
        E_{ni}=E_{ni}^{(0)}+\xi E_{ni}^{(1)}+\xi^2 E_{ni}^{(2)}+...\\
        D_{ni}=D_{ni}^{(0)}+\xi D_{ni}^{(1)}+\xi^2 D_{ni}^{(2)}+...\\
    \end{cases}.
\end{equation}

From the first equation of system \eqref{S3_system_floquet} and system \eqref{S3_peturbation theory} we obtain the following expression

\begin{equation*}
    (\xi^2 k_z^2+2\xi k_z nb+n^2b^2)(E_{nx}^{(0)}+\xi E_{nx}^{(1)}+\xi^2 E_{nx}^{(2)}+\xi^3 E_{nx}^{(3)})-(\xi^2 k_x k_z +\xi k_x nb)\cdot (E_{nz}^{(0)}+\xi E_{nz}^{(1)}+\xi^2 E_{nz}^{(2)}+\xi^3 E_{nz}^{(3)}) = 
\end{equation*}

\begin{equation}
    \label{S3_first eq of system}
    =\xi^2 q^2 (D_{nx}^{(0)}+\xi D_{nx}^{(1)}+\xi^2 D_{nx}^{(2)}+\xi^3 D_{nx}^{(3)}).
\end{equation}

Equating the terms with the same power of $\xi$ results in the following expressions

\begin{equation}
\label{S3_E_nx 0}
    E_{nx}^{(0)}=0,
\end{equation}

\begin{equation}
\label{S3_E_nx 1}
    E_{nx}^{(1)}=\frac{k_x}{nb}E_{nz}^{(0)},
\end{equation}

\begin{equation}
    \label{S3_E_nx 2}
    E_{nx}^{(2)}=-\frac{k_x k_z}{n^2 b^2}E_{nz}^{(0)}+\frac{k_x}{nb}E_{nz}^{(1)}+\frac{q^2}{n^2 b^2} D_{nx}^{(0)},
\end{equation}

\begin{equation}
    \label{S3_E_nx 3}
    E_{nx}^{(3)}=\frac{q^2}{n^2 b^2}D_{nx}^{(1)}+\frac{k_x}{nb}E_{nz}^{(2)}+\frac{k_x k_z}{n^2 b^2} E_{nz}^{(1)}-\frac{2 k_z}{nb}E_{nx}^{(2)}-\frac{k_z^2}{n^2 b^2}E_{nx}^{(1)}.
\end{equation}

Analogous steps for electric displacement components lead to the following equation 

\begin{equation}
    (\xi k_z+nb)(D_{nz}^{(0)}+\xi D_{nz}^{(1)}+\xi^2 D_{nz}^{(2)}+\xi^3 D_{nz}^{(3)})=-\xi k_x (D_{nx}^{(0)}+\xi D_{nx}^{(1)}+\xi^2 D_{nx}^{(2)}+\xi^3 D_{nx}^{(3)}),
\end{equation}

which in turn yields

\begin{equation}
    \label{S3_D zn 0}
    D_{nz}^{(0)}=0,
\end{equation}

\begin{equation}
    \label{S3_D nz 1}
    D_{nz}^{(1)}=-\frac{k_x}{nb}D_{nx}^{(0)},
\end{equation}

\begin{equation}
    \label{S3_D nz 2}
    D_{nz}^{(2)}=-\frac{k_x}{nb}D_{nx}^{(1)}+\frac{k_x k_z}{n^2 b^2}D_{nx}^{(0)},
\end{equation}

\begin{equation}
    \label{S3_D nz 3}
    D_{nz}^{(3)}=-\frac{k_x}{nb}D_{nx}^{(2)}+\frac{k_x k_z}{n^2 b^2}D_{nx}^{(1)}-\frac{k_x k_z^2}{n^3 b^3} D_{nx}^{(0)}.
\end{equation}

By combining the second and the third equations of the systems \eqref{S3_system_floquet} and \eqref{S3_peturbation theory}, we derive 

\begin{minipage}{0.45\textwidth}
  \begin{equation}
        \label{S3_E nz 0}
        E_{n z}^{(0)}= \left(\zeta_{n}\right)_{z i} D_{0 i} ,
    \end{equation}

  \begin{equation}
        \label{S3_E nz 1}
        E_{nz}^{(1)}=\sum_{m \neq 0} (\zeta_{n-m})_{zi} D_{mi}^{(1)},
    \end{equation}

  \begin{equation}
        \label{S3_E nz 2}
        E_{nz}^{(2)}=\sum_{m \neq 0}  (\zeta_{n-m})_{zi} D_{mi}^{(2)},
    \end{equation}
\end{minipage}
\hfill
\begin{minipage}{0.45\textwidth}
    \begin{equation}
  \label{S3_D nx 0}
      D_{n x}^{(0)}= \left(\varepsilon_{n}\right)_{x i} E_{0 i} ,
  \end{equation}

  \begin{equation}
      \label{S3_D nx 1}
      D_{nx}^{(1)}=\sum_{m \neq 0}   (\varepsilon_{n-m})_{xi}E_{mi}^{(1)},
  \end{equation}

   \begin{equation}
      \label{S3_D nx 2}
      D_{nx}^{(2)}=\sum_{m \neq 0}   (\varepsilon_{n-m})_{xi}E_{mi}^{(2)}.
  \end{equation}
\end{minipage}

For the zero Floquet harmonic ($n=0$), which provides the major contribution, from the second and the third equations of system \eqref{S3_system_floquet} we have

\begin{equation}
    \label{S3_system_all_terms}
    \left\{\begin{array}{l}
E_{0 z}= \left(\zeta_{0}\right)_{z i} D_{0 i}+\sum_{n \neq 0} \left(\zeta_{-n}\right)_{z i} \left[D_{n i}^{(0)}+D_{n i}^{(1)}+D_{n i}^{(2)}+D_{n i}^{(3)}\right], \vspace{0.5 cm}\\
D_{0 x}= \left(\varepsilon_{0}\right)_{x i} E_{0 i}+\sum_{n \neq 0}  \left(\varepsilon_{-n}\right)_{x i}\left[E_{n i}^{(0)}+E_{n i}^{(1)}+E_{n i}^{(2)}+E_{n i}^{(3)}\right].
\end{array}\right.
\end{equation}
where we set the expansion parameter to $1$ as in the original equations.


\subsection{Zero and first order of perturbation theory}

To find the zero and the first contributing terms in the system \eqref{S3_system_all_terms}, we analyze the following system of equations

\begin{equation}
    \label{S3_system_0_and_1_terms}
    \left\{\begin{array}{l}
E_{0 z}= \left(\zeta_{0}\right)_{z i} D_{0 i}+\sum_{n \neq 0} \left(\zeta_{-n}\right)_{z i} \left[D_{n i}^{(0)}+ D_{n i}^{(1)}\right]\vspace{0.35 cm}\\
D_{0 x}= \left(\varepsilon_{0}\right)_{x i} E_{0 i}+\sum_{n \neq 0}  \left(\varepsilon_{-n}\right)_{x i} \left[E_{n i}^{(0)} + E_{n i}^{(1)}\right]
\end{array}\right.
\end{equation}

For the first equation of the system \eqref{S3_system_0_and_1_terms}, the terms on the right-hand side with $n\neq 0$ vanish. This follows from the structure of the Fourier components of inverse permittivity  \eqref{S3_non-zero xi comp}, where $(\zeta_1)_{zi}=(\zeta_{-1})_{zi}=0$ for any $i$. It means that there is only one term left
\begin{equation}
    \label{S3_D 0 z 0}
    D_{0 z}=\varepsilon_{3} E_{0 z}.
\end{equation}

The second summand of the second equation of the system \eqref{S3_system_0_and_1_terms} can be rewritten using \eqref{S3_E_nx 0} in the following form
\begin{equation*}
    \label{S3_D0X_0_1}
    \sum_{n \neq 0}  \left(\varepsilon_{-n}\right)_{x i} \left[E_{n i}^{(0)} + E_{n i}^{(1)}\right]=\sum_{n \neq 0}  \left[\left(\varepsilon_{-n}\right)_{x y}E_{n y}^{(0)}+\left(\varepsilon_{-n}\right)_{x z}E_{n z}^{(0)} +\right.
\end{equation*}

\begin{equation}
\label{S3_D0X_0_1}
   \left. +\left(\varepsilon_{-n}\right)_{x x}E_{n x}^{(1)}+\left(\varepsilon_{-n}\right)_{x y}E_{n y}^{(1)}+\left(\varepsilon_{-n}\right)_{x z}E_{n z}^{(1)}\right].
\end{equation}

According to \eqref{S3_non-zero varep comp}, $\left(\varepsilon_{-n}\right)_{x z}=0$ for any $n$. It means that the second and the last term of \eqref{S3_D0X_0_1} vanishes. Terms containing $E_{n y}^{(0)}$ and $E_{n y}^{(1)}$ are negligible too. Indeed, the third equation \eqref{S3_system of eq} yields


\begin{equation}
    \label{S3_E_ny_0_1}
        E_{n y}=\frac{q^{2}}{k_{x}^{2}+\left(k_{z}+n b\right)^{2}} D_{n y}=\frac{q^{2}}{n^{2} b^{2}} \cdot\left(1-\frac{k_{z}^{2}+k_{x}^{2}+2 k_{z} n b}{n^{2} b^{2}}\right) D_{n y}+O\left(\xi^{5}\right).
\end{equation}

The analysis of the equation \eqref{S3_E_ny_0_1} demonstrates that there is no contribution from $E_{n y}^{(0)}$ and $E_{n y}^{(1)}$. Therefore, the second equation of the system \eqref{S3_system_0_and_1_terms} can be represented as 

\begin{equation}
    \label{S3_D0x__}
    D_{0 x}= \left(\varepsilon_{0}\right)_{x i} E_{0 i}+\sum_{n \neq 0}  \left(\left(\varepsilon_{-n}\right)_{x x}E_{n x}^{(1)}\right).
\end{equation}

From equations \eqref{S3_E_nx 1}, \eqref{S3_D nz 1}, \eqref{S3_E nz 0} and \eqref{S3_D nx 0} we obtain

\begin{equation}
    \label{S3_System Enx1_Dnz1}
    \left\{
    \begin{array}{l}
    \displaystyle E_{n x}^{(1)} = \frac{k_{x}}{n b}   \left(\zeta_{n}\right)_{zi} D_{0 i}, \vspace{0.2 cm}\\
    \displaystyle D_{n z}^{(1)} = -\frac{k_{x}}{n b}   \left(\varepsilon_{n}\right)_{x i} E_{0 i}.
    \end{array}
    \right.
\end{equation}

By substituting the first equation of the \eqref{S3_System Enx1_Dnz1} into \eqref{S3_D0x__}, it immediately follows 

\begin{equation}
    \label{S3_0}
    \sum_{n \neq 0}  \left(\varepsilon_{-n}\right)_{x x}E_{n x}^{(1)}=\sum_{n \neq 0} \left(\varepsilon_{-n}\right)_{x x}\frac{k_{x}}{n b}  \left(\zeta_{n}\right)_{zi} D_{0 i}=0.
\end{equation}

\noindent The last equality is valid since $(\zeta_1)_{zi}=(\zeta_{-1})_{zi}=0$. Finally, the second equation of the system \eqref{S3_system_0_and_1_terms} can be written in the following form

\begin{equation}
    \label{D0x_0_1_final}
    D_{0 x}=\bar{\varepsilon}E_{0x}.
\end{equation}

\subsection{Second order of perturbation theory}

The analysis of the second-order corrections to the fields reduces to the consideration of the following sums

\begin{equation}
    \label{S3_second order 1}
    \sum_{n \neq 0}  \left(\varepsilon_{-n}\right)_{x i}E_{n i}^{(2)}=\sum_{n \neq 0}\left[\left(\varepsilon_{-n}\right)_{xx}E_{nx}^{(2)}+\left(\varepsilon_{-n}\right)_{xy}E_{ny}^{(2)}\right],
\end{equation}

\begin{equation}
    \label{S3_second order 2}
    \sum_{n \neq 0}  \left(\zeta_{-n}\right)_{z i}D_{n i}^{(2)}=\sum_{n \neq 0}\left[\left(\zeta_{-n}\right)_{zx}D_{nx}^{(2)}+\left(\zeta_{-n}\right)_{zy}D_{ny}^{(2)}\right].
\end{equation}

Consider the second order of $E_{nx}$. After substituting equations \eqref{S3_E nz 0}, \eqref{S3_E nz 1} and \eqref{S3_D nx 0} into \eqref{S3_E_nx 2}, we obtain the following expression

\begin{equation*}
    \sum_{n\neq 0}\left(\varepsilon_{-n}\right)_{xx}E_{n x}^{(2)}=
\end{equation*}

\begin{equation}
    \label{S3_Enx_2}
    =-\sum_{n\neq 0}\left(\varepsilon_{-n}\right)_{xx}\frac{k_x k_{z}}{n^{2} b^{2}}  \left(\zeta_{n}\right)_{z i} D_{0 i}+\sum_{n\neq 0}\left(\varepsilon_{-n}\right)_{xx}\frac{k_x}{n b} \sum_{m \neq 0}  \left(\zeta_{n-m}\right)_{z i} D_{m i}^{(1)}+\sum_{n\neq 0}\left(\varepsilon_{-n}\right)_{xx}\frac{q^{2}}{n^{2} b^{2}}  \left(\varepsilon_{n}\right)_{x i} E_{0 i}.
\end{equation}

 The first term of equation \eqref{S3_Enx_2} equals to zero because of $(\zeta_{\pm 1})_{zi}=0$: 

\begin{equation*}
    -\sum_{n\neq 0}\left(\varepsilon_{-n}\right)_{xx}\frac{k_x k_{z}}{n^{2} b^{2}}  \left(\zeta_{n}\right)_{z i} D_{0 i}=0.
\end{equation*}

There are only terms left on the second summand of \eqref{S3_Enx_2}, when $n=m$: in this case $(\zeta_0)_{zz}\neq 0$. Taking into account the second expression of the system \eqref{S3_System Enx1_Dnz1}, it is obtained

\begin{equation*}
         \sum_{n\neq 0}\left(\varepsilon_{-n}\right)_{xx}\frac{k_x}{n b} \sum_{m \neq 0}  \left(\zeta_{n-m}\right)_{z i} D_{m i}^{(1)}= -\sum_{n \neq 0}\left(\varepsilon_{-n}\right)_{x x} \frac{k_{x}^{2}}{n^{2} b^{2}}\left(\zeta_{0}\right)_{z z} \left(\varepsilon_{n}\right)_{x i} E_{0 i}=
 \end{equation*}

\begin{equation}
\label{S3_Enx_2_2}
      = -\sum_{n \neq 0}\left(\varepsilon_{-n}\right)_{x x} \frac{k_{x}^{2}}{n^{2} b^{2}} \frac{1}{\varepsilon_{3}}\left(\left(\varepsilon_{n}\right)_{x x} E_{0 x}+\left(\varepsilon_{n}\right)_{x y} E_{0 y}\right)= -\frac{1}{2\varepsilon_3}\left(\frac{k_{x}\Delta \varepsilon}{ b} \right)^{2} E_{0 x}.
\end{equation}

 The last term of the equation \eqref{S3_Enx_2} can be written explicitly as

 \begin{equation}
\label{S3_Enx_2_3}
     \sum_{n \neq 0}\left(\varepsilon_{-n}\right)_{x x} \frac{q^{2}}{n^{2} b^{2}}  \left(\varepsilon_{n}\right)_{x i} E_{0 i}=\sum_{n \neq 0}\left(\varepsilon_{n}\right)_{x x} \frac{q^{2}}{n^{2} b^{2}} \left(\left(\varepsilon_{n}\right)_{x x} E_{0 x}+\left(\varepsilon_{n}\right)_{x y} E_{0 y}\right)=\frac{1}{2}\left(\frac{q\Delta \varepsilon}{b} \right)^{2}E_{0 x}.
 \end{equation}

Consider the second order of $E_{ny}$. By extracting the quadratic part on the small parameter from equation \eqref{S3_E_ny_0_1}, we obtain

\begin{equation}
    \label{S3_E_ny_1}
    E_{ny}=\frac{q^2}{n^2 b^2}D_{ny}.
\end{equation}

Taking into account the connection between electric vector and electric displacement vector 

\begin{equation*}
    D_{n k}=\sum_{m}  \left(\varepsilon_{n-m}\right)_{k i} E_{m i},
\end{equation*}

it is possible to write 

\begin{equation}
\label{S3_D_ny_1}
    D_{n y}=\sum_{m}  \left(\varepsilon_{n-m}\right)_{y i}E_{mi}= (\varepsilon_{n})_{yi}E_{0i}.
\end{equation}

After substituting \eqref{S3_D_ny_1} to \eqref{S3_E_ny_1}, we have

\begin{equation}
    \label{S3_E_ny_2}
    E_{n y}^{(2)}=\frac{q^{2}}{n^{2} b^{2}}  \left(\varepsilon_{n}\right)_{y i} E_{0 i}+O\left(\xi^{3}\right).
\end{equation}

Finally, we can plug \eqref{S3_E_ny_2} into \eqref{S3_second order 1} and get

\begin{equation}
\label{S3_E_ny_2_34}
    \sum_{n \neq 0}\left(\varepsilon_{-n}\right)_{x y} \cdot \frac{q^{2}}{n^{2} b^{2}}  \left(\varepsilon_{n}\right)_{y i} E_{0 i}= \sum_{n \neq 0}\left(\varepsilon_{-n}\right)_{x y} \cdot \frac{q^{2}}{n^{2} b^{2}}\left(\left(\varepsilon_{n}\right)_{y x} E_{0 x}+\left(\varepsilon_{n}\right)_{y y} E_{0 y}\right)=\frac{1}{2}\left(\frac{q\Delta \varepsilon}{b} \right)^{2}E_{0 x}.
\end{equation}

Therefore, equation \eqref{S3_second order 1} can be written as

\begin{equation}
    \label{S3_second order E fin}
    \sum_{n \neq 0}  \left(\varepsilon_{-n}\right)_{x i}E_{n i}^{(2)}=\frac{\Delta \varepsilon^2}{2b^2}\left(2q^2-\frac{k_x^2}{\varepsilon_3}\right).
\end{equation}

The sum \eqref{S3_second order 2} vanishes due to $(\zeta_1)_{zi}=(\zeta_{-1})_{zi}=0$.

Combining the components of the zero, the first and second order fields, we obtain

\begin{equation}
\label{S3_2nd order D0z and D0x}
    \left\{
    \begin{array}{l}
    D_{0 z}=\varepsilon_{3} E_{0 z}, \vspace{0.2 cm}\\
   D_{0 x}=\left[\bar{\varepsilon}+\frac{\Delta \varepsilon^{2}}{2 b^{2}}\left(2 q^{2}-\frac{k_{x}^{2}}{\varepsilon_{3}}\right)\right] E_{0 x}.
    \end{array}
    \right.
\end{equation}

To get effective parameters in the case of oblique incidence, it is necessary to know all components of the electric displacement vector. In order to find $D_{0y}$, we consider equation \eqref{S3_D_ny_1}:

\begin{equation}
    \label{S3_D_0y_}
    D_{0y}=\sum_{m} (\varepsilon_{-m})_{y i}E_{m i}=\underbrace{ (\varepsilon_{0})_{y i}E_{0i}}_{\bar{\varepsilon}E_{0y}}+\sum_{n\neq 0} (\varepsilon_{-n})_{y i}E_{ni}^{(2)}.
\end{equation}

Consider separately the second term of the equation \eqref{S3_D_0y_}:

\begin{equation}
    \label{S3_D_0y_2}
    \sum_{n\neq 0} (\varepsilon_{-n})_{y i}E_{ni}^{(2)}=\sum_{n\neq 0}\left[(\varepsilon_{-n})_{y x}E_{nx}^{(2)}+(\varepsilon_{-n})_{y y}E_{ny}^{(2)}\right].
\end{equation}

Similar to expression \eqref{S3_Enx_2} for the $E_{nx}^{(2)}$ and using equation \eqref{S3_E_ny_2}, it can be obtained 

\begin{equation*}
         \sum_{n\neq 0} (\varepsilon_{-n})_{y i}E_{ni}^{(2)}=-\sum_{n\neq 0}\left(\varepsilon_{-n}\right)_{y x} \frac{k_x k_z}{n^{2} b^{2}} \left(\zeta_{n}\right)_{z i} D_{0 i}+
        +\sum_{n \neq 0}\left(\varepsilon_{-n}\right)_{y x} \frac{k_x}{n b} \sum_{m \neq 0}  \left(\zeta_{n-m}\right)_{z i} D_{m i}^{(1)}+
\end{equation*}
\begin{equation}
    +\sum_{n \neq 0}\left(\varepsilon_{-n}\right)_{y x} \frac{q^{2}}{n^{2} b^{2}}  \left(\varepsilon_{n}\right)_{x i} E_{0 i}+\sum_{n \neq 0}\left(\varepsilon_{-n}\right)_{y y} \frac{q^{2}}{n^{2} b^{2}}  \left(\varepsilon_{n}\right)_{y i} E_{0 i}=\frac{\Delta\varepsilon^2}{2b^2}\left(q^2 - \frac{k_x^2}{\varepsilon_3}\right)E_{0y}
\end{equation}

Therefore, the $D_{0y}$ component with accuracy up to the second order of small parameter has the following form

\begin{equation}
    \label{S3_D_0y_4}
    D_{0 y}=\left[\bar{\varepsilon}+\frac{\Delta \varepsilon^{2}}{2 b^{2}}\left(2 q^{2}-\frac{k_{x}^{2}}{\varepsilon_{3}}\right)\right] E_{0 y}
\end{equation}

Merging \eqref{S3_2nd order D0z and D0x} and \eqref{S3_D_0y_4} gives

\begin{equation}
\label{S3_2nd order D0z and D0x}
    \left\{
    \begin{array}{l}
       D_{0 x}=\left[\bar{\varepsilon}+\frac{\Delta \varepsilon^{2}}{2 b^{2}}\left(2 q^{2}-\frac{k_{x}^{2}}{\varepsilon_{3}}\right)\right] E_{0 x}, , \vspace{0.2 cm}\\
    D_{0 y}=\left[\bar{\varepsilon}+\frac{\Delta \varepsilon^{2}}{2 b^{2}}\left(2 q^{2}-\frac{k_{x}^{2}}{\varepsilon_{3}}\right)\right] E_{0 y}, \vspace{0.2 cm}\\
    D_{0 z}=\varepsilon_{3} E_{0 z}.
    \end{array}
    \right.
\end{equation}

\subsection{Third order of perturbation theory}

However, to capture the effective chirality, one has to consider the third order of perturbation theory. We consider the following sum

\begin{equation}
    \label{S3_3rd order}
        \sum_{n\neq0} (\varepsilon_{-n})_{xi}E_{ni}^{(3)}=\sum_{n \neq 0}\left(\varepsilon_{-n}\right)_{x x} E_{n x}^{(3)}+\sum_{n \neq 0}\left(\varepsilon_{-n}\right)_{x y} E_{n y}^{(3)}\\
\end{equation}

To calculate the sum with $E_{nx}^{(3)}$, it is necessary to consider the equation \eqref{S3_E_nx 3}.

First term in this equation vanishes:

\begin{equation*}
    D^{(1)}_{nx} = \sum_{m\neq0} \sum_{i=1}^3 (\varepsilon_{n-m})_{xi}E_{mi}^{(1)} = \sum_{m\neq0}(\varepsilon_{n-m})_{xx}E_{mx}^{(1)} = \sum_{m\neq 0}(\varepsilon_{n-m})_{xx}\frac{k_x}{mb}E_{mz}^{(0)} =
\end{equation*}
\begin{equation}
    \label{S3_Enx3_1}
    = \sum_{m\neq 0}(\varepsilon_{n-m})_{xx}\frac{k_x}{mb} (\zeta_{m})_{zi}D_{0i} = 0,
\end{equation}

\noindent where it was taken into account that $E_{my}^{(1)} = 0$ and $\zeta_{\pm1} =0$. The term with $E_{mz}^{(1)}$ also vanishes due to $(\varepsilon_{n-m})_{xz} = 0$.

The second term in equation \eqref{S3_E_nx 3} when substituted into equation \eqref{S3_3rd order} reduces to 

\begin{equation}
\label{S3_Enx3_2_1}
    \sum_{n\neq0} (\varepsilon_{-n})_{xx}\frac{k_x}{nb}\sum_{m\neq0} (\zeta_{n-m})_{zi}D_{mi}^{(2)},
\end{equation}

\noindent where equation \eqref{S3_E nz 2} was used.

It is obtained non-zero value if $n=m$ and $i=z$:

\begin{equation}
    \label{S3_Enx3_2_2}
        \sum_{n\neq0} (\varepsilon_{-n})_{xx}\frac{k_x}{nb}(\zeta_{0})_{zz}D_{nz}^{(2)}.
\end{equation}

Using equation \eqref{S3_D nz 2} we get 2 sums. We examine the first one:

\begin{equation*}
-\sum_{n\neq0}(\varepsilon_{-n})_{xx}\frac{k_x^2}{n^2b^2}\frac{1}{\varepsilon_3}D_{nx}^{(1)} = -\frac{1}{\varepsilon_3}\sum_{n\neq0}(\varepsilon_{-n})_{xx}\frac{k_x^2}{n^2b^2}\sum_{m \neq 0}\left[(\varepsilon_{n-m})_{xx}E_{mx}^{(1)} + (\varepsilon_{n-m})_{xy}E_{my}^{(1)}\right] = 
\end{equation*}

\begin{equation}
\label{S3_Enx3_2_3}
    =-\frac{1}{\varepsilon_3}\sum_{n\neq0}(\varepsilon_{-n})_{xx}\frac{k_x^2}{n^2b^2}\sum_{m \neq 0}(\varepsilon_{n-m})_{xx}\frac{k_x}{mb} (\zeta_{m})_{zi}D_{0i} = 0,
\end{equation}

\noindent where the last equality is valid due to $(\zeta_{\pm 1})_{zi} = 0$ for any $i$.

The second one sum reduces to

\begin{equation*}
    \sum_{n\neq0}(\varepsilon_{-n})_{xx}\frac{k_x^2 k_z}{n^3 b^3}\frac{1}{\varepsilon_3} (\varepsilon_n)_{xi}E_{0i} = \sum_{n\neq0}(\varepsilon_{-n})_{xx}\frac{k_x^2 k_z}{n^3 b^3}\frac{1}{\varepsilon_3}\left[(\varepsilon_n)_{xx}E_{0x}+(\varepsilon_n)_{xy}E_{0y}\right] = 
\end{equation*}
\begin{equation}
\label{S3_Enx3_2_4}
    = -i\frac{\Delta\varepsilon^2}{2}\frac{k_x^2 k_z}{\varepsilon_3 b^3}E_{0y}.
\end{equation}

The third term in equation \eqref{S3_E_nx 3} when substituted into equation \eqref{S3_3rd order} reduces to 

\begin{equation*}
     \sum_{n\neq0}(\varepsilon_{-n})_{xx}\frac{k_x k_z}{n^2 b^2}\sum_{m\neq0} (\zeta_{n-m})_{zi}D_{mi}^{(1)} = \sum_{n\neq0}(\varepsilon_{-n})_{xx}\frac{k_x^2 k_z}{n^3 b^3} \frac{1}{\varepsilon_3} (\varepsilon_n)_{xi}E_{0i} = 
\end{equation*}
\begin{equation}
\label{S3_Enx3_2_5}
    =i\frac{\Delta\varepsilon^2}{2}\frac{k_x^2 k_z}{\varepsilon_3 b^3}E_{0y}.
\end{equation}

The forth term in equation \eqref{S3_E_nx 3} when substituted into equation \eqref{S3_3rd order} reduces to 

\begin{equation}
    \label{S3_Enx3_2_6}
    -\sum_{n\neq0}(\varepsilon_{-n})_{xx}\frac{2k_z}{nb}E_{nx}^{(2)}.
\end{equation}

Here we need to substitute the equation \eqref{S3_E_nx 2} and take into account expressions \eqref{S3_E nz 1} and \eqref{S3_D nx 0} as follows

\begin{equation}
    \label{S3_Enx3_2_7}
    E_{nx}^{(2)} = - \frac{k_x k_z}{n^2 b^2}  (\zeta_{n})_{zi}D_{0i} + \frac{k_x}{nb}\sum_{m\neq0} (\zeta_{n-m})_{zi}D_{mi}^{(1)} + \frac{q^2}{n^2 b^2} (\varepsilon_{n})_{xi}E_{0i}.
\end{equation}

After substituting the expression \eqref{S3_Enx3_2_7} into equation \eqref{S3_Enx3_2_6}, the first term will vanish. The second term can be written as

\begin{equation*}
    -\sum_{n\neq0} (\varepsilon_{-n})_{xx}\frac{2k_z k_x}{n^2 b^2}\sum_{m\neq0}   (\zeta_{n-m})_{zi}D_{mi}^{(1)} = \sum_{n\neq0} (\varepsilon_{-n})_{xx}\frac{2k_z k_x^2}{n^3 b^3}\frac{1}{\varepsilon_3}\left[(\varepsilon_n)_{xx}E_{0x} + (\varepsilon_n)_{xy}E_{0y}\right] = 
\end{equation*}

\begin{equation}
    \label{S3_Enx3_2_8}
    = -i\Delta\varepsilon^2
\frac{k_x^2 k_z}{\varepsilon_3 b^3}E_{0y}.
\end{equation}

The third term reduces to 

\begin{equation}
\label{S3_Enx3_2_9}
    -\sum_{n\neq0}(\varepsilon_{-n})_{xx}\frac{2k_z q^2}{n^3b^3} (\varepsilon_{n})_{xi}E_{0i} = i\Delta\varepsilon^2\frac{k_z q^2}{b^3}E_{0y}.
\end{equation}

The fifth term in equation \eqref{S3_E_nx 3} when substituted into equation \eqref{S3_3rd order} reduces to zero due to $(\zeta_{\pm 1})_{zi} = 0$ for any $i$.

Therefore, the first part for the equation \eqref{S3_3rd order} can be written as

\begin{equation}
    \label{S3_Enx3_final}
    \sum_{n\neq0}(\varepsilon_{-n})_{xx}E_{nx}^{(3)} = i\Delta\varepsilon^2 \frac{k_z}{b^3}\left(q^2 - \frac{k_x^2}{\varepsilon_3}\right)
\end{equation}

To get the second part, consisting $E_{ny}^{(3)}$, it is necessary to return to the equation \eqref{S3_E_ny_0_1} and select the component proportional to $\xi^3$. Substituting there the expression \eqref{S3_D_ny_1}, we obtain the following equation

\begin{equation}
\label{S3_Eny3__}
    E_{ny}^{(3)} = -\frac{2k_z q^2}{n^3 b^3} \sum_{i=1}^3(\varepsilon_n)_{yi}E_{0i}.
\end{equation}

Substituting it into sum leads us to the following expression

\begin{equation}
    \label{S3_Eny3_final}
    \sum_{n\neq0}(\varepsilon_{-n})_{xy}E_{ny}^{(3)}= i\Delta\varepsilon^2\frac{k_z q^2}{b^3}E_{0y}
\end{equation}

Therefore, $D_{0x}$ with accuracy up to the third order of small parameter has the following form

\begin{equation}
    \label{S3_D_0X_final}
     D_{0 x}=\left[\bar{\varepsilon}+\frac{\Delta \varepsilon^{2}}{2 b^{2}}\left(2 q^{2}+\frac{k_{x}^{2}}{\varepsilon_{3}}\right)\right] E_{0 x} + i\frac{\Delta\varepsilon^2 k_z}{b^3}\left(2q^2 - \frac{k_x^2}{\varepsilon_3}\right) E_{0y}.
\end{equation}

In a similar way, $D_{0y}$ can be found as

\begin{equation}
    \label{S3_D0Y_3_1}
    D_{0y} =  (\varepsilon_0)_{yi}E_{oi} + \sum_{n\neq0} (\varepsilon_{-n})_{yi}E_{ni}^{(3)}.
\end{equation}

The first summation is reduced to $\bar{\varepsilon}E_{0y}$. In the second summation it is necessary to do the same calculations but replace $(\varepsilon_{-n})_{xi}$ on $(\varepsilon_{-n})_{yi}$. Therefore, 
the equation for $D_{oy}$ with accuracy up to the third order of small parameter has the following form

\begin{equation}
    \label{S3_D_0Y_final}
     D_{0 y}=\left[\bar{\varepsilon}+\frac{\Delta \varepsilon^{2}}{2 b^{2}}\left(2 q^{2}-\frac{k_{x}^{2}}{\varepsilon_{3}}\right)\right] E_{0 y} - i\frac{\Delta\varepsilon^2 k_z}{b^3}\left(2q^2 - \frac{k_x^2}{\varepsilon_3}\right) E_{0y}.   
\end{equation}

Finally, the connection between electric displacement vector and the electric field with accuracy up to the third order of a small parameter $a/\lambda$ for the structure consisting of anisotropic infinitely thin layers with continuously rotating anisotropy axis has the form

\begin{equation}
\label{S3_final system}
    \left\{
    \begin{array}{l}
       D_{0 x}=\left[\bar{\varepsilon}+\frac{\Delta \varepsilon^{2}}{2 b^{2}}\left(2 q^{2}-\frac{k_{x}^{2}}{\varepsilon_{3}}\right)\right] E_{0 x} + i\frac{\Delta\varepsilon^2 k_z}{b^3}\left(2q^2 - \frac{k_x^2}{\varepsilon_3}\right) E_{0y},  \vspace{0.3 cm}\\
    D_{0 y}=\left[\bar{\varepsilon}+\frac{\Delta \varepsilon^{2}}{2 b^{2}}\left(2 q^{2}-\frac{k_{x}^{2}}{\varepsilon_{3}}\right)\right] E_{0 y} - i\frac{\Delta\varepsilon^2 k_z}{b^3}\left(2q^2 - \frac{k_x^2}{\varepsilon_3}\right) E_{0x}, \vspace{0.3 cm}\\
    D_{0 z}=\varepsilon_{3} E_{0 z}.
    \end{array}
    \right.
\end{equation}

\subsection{Reducing the material equations to canonical bianisotropic form}

Equations~\eqref{S3_final system} provide the constitutive relations for the averaged fields in the nonlocal (spatially dispersive) framework. To convert these to the more convenient bianisotropic form, we redefine the auxiliary fields ${\bf H}$ and ${\bf D}$ such that Maxwell's equations are kept invariant.

To that end, we denote some factors in the \eqref{S3_final system} for simplicity as follows

\begin{equation}
\label{S3_A}
    A = \bar{\varepsilon}+\frac{\Delta \varepsilon^{2}}{2 b^{2}}\left(2 q^{2}-\frac{k_{x}^{2}}{\varepsilon_{3}}\right),
\end{equation}
\begin{equation}
\label{S3_B}
    B = \frac{\Delta\varepsilon^2}{b^3}\left(2q^2 - \frac{k_x^2}{\varepsilon_3}\right).
\end{equation}

Hence, the system of equations \eqref{S3_final system} can be rewritten as follows

\begin{equation}
\label{S3_system with A B}
    \left\{
    \begin{array}{l}
       D_{0 x}=A E_{0 x} + ik_z B E_{0y},  \vspace{0.1 cm}\\
    D_{0 y}= -ik_z B E_{0x} + A E_{0y}, \vspace{0.1 cm}\\
    D_{0 z}=\varepsilon_{3} E_{0 z}
    \end{array}
    \right.
\end{equation}

or

\begin{equation}
    \label{S3_D0 2}
    \v{D}_0  = \hat{\varepsilon}_{\text{eff}}\v{E}_0 - i B k_z \hat{e}_z^{\times}\v{E}_0
\end{equation}

\noindent where 

\begin{equation}
    \label{S3_D0_A_B}    \hat{\varepsilon}_{\text{eff}} = \begin{pmatrix}
        A & 0& 0\\
        0 & A& 0\\
        0 & 0& \varepsilon_3
    \end{pmatrix} \qquad \text{and} \qquad \hat{e}_z^{\times} = \begin{pmatrix}
        0 & -1& 0\\
        1 & 0& 0\\
        0 & 0& 0
    \end{pmatrix}.
\end{equation}

Similar to normal incidence case, to reduce the material equations to canonical bianisotropic form, it is necessary to redefine magnetic field $\v{H}_0$ as follows

\begin{equation}
    \label{S3_H0}
    \v{H}_0 = \v{B}_0 - i\hat{\kappa}\v{E}_0.
\end{equation}

Consider the following Maxwell's equation for the monochromatic plane wave

\begin{equation}
    \label{S3_Max eq}
    \v{k}\times \v{H}_0 = -q \v{D}_0 '.
\end{equation}

Expressing $\v{D}_0'$ from the equation \eqref{S3_Max eq} and substituting there vector $\v{H}_0$ lead us to the following redefinition of electric displacement:

\begin{equation}
    \label{S3_D0'}
    \v{D}_0' = \v{D}_0 + \frac{i}{q}\hat{k}^{\times}\hat{\kappa}\v{E}_0,
\end{equation}

\noindent where 

\begin{equation}
    \label{S3_kx}
    \hat{k}^{\times} = \begin{pmatrix}
        0 & -k_z& 0\\
        k_z & 0& -k_x\\
        0 & k_x& 0
    \end{pmatrix}.
\end{equation}

After substituting \eqref{S3_D0 2} to \eqref{S3_D0'}, we obtain

\begin{equation}
    \label{S3_D0'_2}
    \v{D}_0' = \hat{\varepsilon}_{\text{eff}}\v{E}_0 - iB k_z \hat{e}_z^{\times} \v{E}_0 + \frac{i}{q}\hat{k}^{\times}\hat{\kappa}\v{E}_0.
\end{equation}

On the other hand, to obtain the conventional material equations with chirality tensor, the following equation should be fulfilled:
\begin{equation}
    \label{S3_demand}
    - iB k_z \hat{e}_z^{\times} \v{E}_0 + \frac{i}{q}\hat{k}^{\times}\hat{\kappa}\v{E}_0 = i \hat{\kappa}\v{H}_0.
\end{equation}

From equations \eqref{S3_H0} and \eqref{S3_demand} it follows

\begin{equation}
    \label{S3_demand2}
     - B k_z \hat{e}_z^{\times} \v{E}_0 + \frac{1}{q}\hat{k}^{\times}\hat{\kappa}\v{E}_0 =  \hat{\kappa}\v{B}_0
\end{equation}

\noindent or by applying Maxwell's equation for monochromatic plane wave

\begin{equation}
    \label{S3_demand3}
\hat{k}^{\times}\hat{\kappa}-\hat{\kappa}\hat{k}^{\times} = qBk_z \hat{e}_z^{\times}
\end{equation}

\noindent where the $\hat{\kappa}^2$ term is neglected due to its smallness. 

This is the matrix equation on $\hat{\kappa}$. The solution of matrix equation \eqref{S3_demand3} has the following form

\begin{equation}
    \label{S3_kappa_eff}
    \hat{\kappa}_{\text{eff}} = -\frac{1}{2} \begin{pmatrix}
        qB&0&0\\
        0&qB&0\\
        0&0&-qB
    \end{pmatrix}.
\end{equation}

Hence, the  effective parameters comprising bianisotropic constitutive relations for the general direction of the wave vector have the following form

\begin{equation}
    \label{S3_final_eff_params_eps}
    \hat{\varepsilon}_{\text{eff}} = \begin{pmatrix}
        \bar{\varepsilon}+\frac{\Delta \varepsilon^{2}}{2 b^{2}}\left(2 q^{2}-\frac{k_{x}^{2}}{\varepsilon_{3}}\right)&0&0\\
        0&\bar{\varepsilon}+\frac{\Delta \varepsilon^{2}}{2 b^{2}}\left(2 q^{2}-\frac{k_{x}^{2}}{\varepsilon_{3}}\right)&0\\
        0&0&\varepsilon_3
    \end{pmatrix},
\end{equation}

\begin{equation}
    \label{S3_final_eff_params_kappa}
    \hat{\kappa}_{\text{eff}} = -\frac{q}{2} \frac{\Delta\varepsilon^2}{b^3}\left(2q^2 - \frac{k_x^2}{\varepsilon_3}\right)\begin{pmatrix}
        1&0&0\\
        0&1&0\\
        0&0&-1
    \end{pmatrix}.
\end{equation}

\bibliography{ChiralLib}